\documentclass{aa}  

\usepackage{graphicx}
\usepackage{multirow}

\usepackage{gensymb}

\usepackage{color}
\usepackage{here}

\newcommand{\ii}{{{\rm i}}}

\usepackage[colorlinks=true,citecolor=blue]{hyperref}
\usepackage{arydshln}
\usepackage{natbib}

\usepackage{comment}

\usepackage{booktabs}

\usepackage{adjustbox}
\usepackage{academicons}
\usepackage{orcidlink}

\usepackage{ulem}

\defcitealias{hotta2025}{H25}

\usepackage{scalerel}
\usepackage{stackengine,wasysym}

\usepackage{txfonts}
\begin{document} 
   \title{
   Inertial modes in a turbulent rotating magneto-convection simulation with the solar-like differential rotation
   }

   \titlerunning{Inertial modes in a rotating magneto-convection simulation with solar-like differential rotation}

   \author{Yuto Bekki \inst{1}\orcidlink{0000-0002-5990-013X}
          \and
          Hideyuki Hotta \inst{2}\orcidlink{0000-0002-6312-7944} 
          \and
          Laurent Gizon\inst{1,3,4}\orcidlink{0000-0001-7696-8665}
          }

          \authorrunning{Bekki et al.}

   \institute{Max-Planck-Institut f{\"u}r Sonnensystemforschung,
              Justus-von-Liebig-Weg 3, 37077 G{\"o}ttingen, Germany\\
              \email{\href{mailto:bekki@mps.mpg.de}{bekki@mps.mpg.de}}
        \and
        Institute for Space-Earth Environmental Research, Nagoya University, Furo-cho, Chikusa-ku, Nagoya, Aichi 464-8601, Japan
        \and
         Institut f{\"u}r Astrophysik, Georg-August-Universit{\"a}t G{\"o}ttingen, Friedrich-Hund-Platz 1, 37077 G{\"o}ttingen, Germany
         \and
         Center for Astrophysics and Space Science, NYUAD Institute, New York University Abu Dhabi, Abu Dhabi, UAE
        }

   \date{Received <-->; accepted <-->}


  \abstract
{
Recent observations have revealed a variety of solar inertial modes, but their nonlinear properties remain largely unexplored in realistic first-principles simulations.
} 
{
We aim to identify and characterize inertial modes in a state-of-the-art nonlinear simulation of rotating magneto-convection, in which solar-like differential rotation was self-consistently reproduced with the solar parameters. 
}
{
We analyzed nine years of high-cadence shell-slice data from the \citet{hotta2025} simulation. 
The spatial eigenfunctions of various low-frequency inertial modes were extracted from the simulation using singular value decomposition.
}
{
We identified major classes of retrograde inertial modes in the simulation, including equatorial Rossby modes, high-latitude modes, and high-frequency retrograde modes, with frequencies and near-surface amplitudes roughly consistent with the solar observations.
In particular, thanks to the strongly baroclinic background achieved in the simulation, the observed large amplitude ($\sim 10$~m~s$^{-1}$) and the characteristic polar spiral pattern of the high-latitude mode at azimuthal order $m=1$ are successfully reproduced.
This indicates that high-latitude modes are excited by baroclinic instability, which can occur even amid turbulent convection.
We further show that the prograde columnar modes undergo a qualitative transition at $m \simeq 6$, separating the fast-propagating convective modes at $m \gtrsim 6$ from thermally diffusive slow modes at $m \lesssim 6$ whose frequencies closely match recent observational inferences of thermal Rossby modes.
Finally, we find that turbulent magnetic fields primarily influence the inertial modes indirectly through modifications of the background state, whereas  their direct effects on the modes are comparatively weak.
}
{
Our results suggest that the observed properties of solar inertial modes are broadly consistent with a simulation in which solar-like differential rotation is maintained in a magnetically dominated regime, although the simulation still overproduces prograde columnar modes at large $m$.
}


   \maketitle
%

\section{Introduction} \label{sec:intro}

Recent observations of horizontal flows at the surface of the Sun have revealed a variety of inertial modes of oscillation \citep[e.g.,][]{gizon2024iau}. 
They are low-frequency modes of global oscillations for which the Coriolis force acts as the primary restoring force. 
Among the modes detected so far are equatorial Rossby modes with radial vorticity that is symmetric across the equator \citep[][]{loeptien2018}; high-frequency retrograde (HFR) modes with antisymmetric radial vorticity \citep[][]{hanson2022}; and high-latitude modes with characteristic spiral patterns around the poles \citep[][]{gizon2021}. 
All of these modes propagate in the retrograde direction in the Carrington rotating frame. 
Although not yet firmly established, tentative detection of prograde-propagating inertial modes has also been reported recently \citep[][]{hanasoge2026,mandal2026}.
These solar inertial modes may not only provide new diagnostics of the solar interior, but also play an important role in the large-scale dynamics of the convection zone \citep[e.g.,][]{bekki2026_review}.

Studies of solar inertial modes have so far relied primarily on linear models \citep[][]{bekki2022a,triana2022,bhattacharya2023, mukhopadhyay2025}.
In these models, the linearized hydrodynamic equations in the Sun's differentially rotating convection zone are solved as an eigenvalue problem, enabling the determination of mode frequencies and eigenfunctions. 
It has been shown that realistic modeling of solar inertial modes requires incorporating the solar background stratification \citep[][]{mukhopadhyay2025}, differential rotation \citep[][]{gizon2020,fournier2022,dey2026}, and latitudinal entropy gradient \citep[][]{bekki2022a}.
The linear models are useful for identifying observed modes and for constraining poorly known properties of the Sun's convection zone, such as the turbulent viscosity, superadiabaticity, and the differential rotation profile at high latitudes \citep[][]{gizon2021,bekki2024a,dey2026}.

Nonlinear models are necessary for investigating the excitation mechanisms and amplitudes of solar inertial modes.
Using simplified two-dimensional (2D) spherical surface models, previous studies have investigated the stochastic excitation of linearly stable equatorial Rossby modes by turbulent convection \citep{philidet2023}, as well as the saturation of linearly unstable high-latitude modes within the framework of weakly nonlinear theory \citep{mushtaq2026}.
Another approach is to artificially impose the solar differential rotation without explicitly solving turbulent convection, and to examine the nonlinear evolution of mode perturbations in the convection zone \citep[][]{souzagomes2026}.
In particular, within this non-convecting framework, \citet{bekki2024b} have successfully demonstrated that high-latitude modes can be baroclinically excited and reach the velocity amplitudes comparable to the observed values.
It was further shown that their nonlinear saturation involves equatorward heat transport, thereby regulating the background baroclinicity and the differential rotation.

The nonlinear models described above are essentially reduced models, in which the background differential rotation is prescribed in an ad-hoc manner. 
A more direct approach is to analyze inertial modes in fully nonlinear rotating convection simulations, where differential rotation and mean entropy gradients are self-consistently generated as a consequence of the resolved dynamics \citep[][]{bekki2022b,blume2024,fuentes2026}.
However, a major difficulty exists: it is extremely challenging to reproduce the observed differential rotation in such first-principles simulations \citep[e.g.,][]{hotta2023_review,kapyla2023_review}. 
To address this issue, previous studies have employed simulations with smaller luminosity \citep{bekki2022b} or faster rotation rate \citep{blume2024} than those of the Sun.
In these simulations, equatorial Rossby modes, HFR modes, and mixed modes between prograde and retrograde inertial modes \citep[as predicted by][]{bekki2022a} have been identified. 
However, the observed polar spiral structure and substantial velocity amplitudes of the high-latitude modes have not been reproduced.
This is likely because both the differential rotation and the latitudinal entropy gradient are much weaker in the simulation than in the Sun.
Furthermore, in these simulations, the large-scale flow patterns tend to be dominated by prograde-propagating columnar convection cells, particularly when rotational influences are strong (characterized by a low Rossby number Ro) \citep[e.g.,][]{miesch2000,miesch2008,hindman2020}. 
Such convective columns are closely related to thermal Rossby modes predicted by linear theories of rotating convection \citep[e.g.,][]{busse1970,gilman1975,busse2002}. 
In this paper, however, we use the broader term prograde columnar modes, as such inertial modes are not necessarily driven by convective instability \citep[][]{glatzmaier1981,bekki2022a,mukhopadhyay2025}.
Despite the dominance in numerical models, observational evidence for such modes in the Sun remains limited.
This discrepancy is crucial because these modes are expected to play an important role in transporting angular momentum and maintaining solar-like differential rotation \citep[e.g.,][]{gilman1977,brun2002,bekki2022b,mori2023a,kapyla2023}. 
These findings collectively imply that a theoretical understanding of solar inertial modes in the fully nonlinear regime cannot be disentangled from the physics responsible for the generation and maintenance of differential rotation.

\citet{hotta2021} performed high-resolution first-principles simulations of rotating magneto-convection at the solar parameters (e.g., rotation rate and luminosity), and demonstrated that solar-like differential rotation can be reproduced in a strongly turbulent regime with high enough fluid and magnetic Reynolds numbers. 
In their simulations, the solar-like differential rotation does not arise from angular momentum transport by columnar convection in a strongly rotationally constrained regime. 
Instead, it is maintained by a vigorous turbulent dynamo action that generates super-equipartition magnetic fields whose Maxwell stresses transport angular momentum radially outward, counteracting the inward transport by Reynolds stresses \citep[][]{hotta2022}.
More recently, \citet[][hereafter \citetalias{hotta2025}]{hotta2025} further performed an even higher-resolution simulation with extended numerical domain and demonstrated that a solar-like near-surface shear layer (NSSL) can also be reproduced. 
In these simulations, the differential rotation exhibits a significant deviation from the Taylor-Proudman state, as inferred by helioseismic observations \citep[e.g.,][]{thompson2003}.
This deviation is largely sustained via thermal wind balance by a latitudinal entropy gradient that arises self-consistently from anisotropic heat transport by strongly magnetized convection \citep[][]{hotta2018}.

This new simulation dataset with high-fidelity differential rotation by \citetalias{hotta2025} provides a useful foundation for the study of solar inertial modes.
It enables us to examine the validity of linear models, the mode excitation and nonlinear saturation mechanisms, and the role of turbulent convection and magnetic fields. 
In this study, we perform a systematic analysis of inertial modes in the \citetalias{hotta2025} simulation. 
We identify and characterize not only the equatorial Rossby modes, HFR modes, and high-latitude modes (observed on the Sun), but also prograde columnar modes, which may be related to recent observations by \citet[][]{hanasoge2026} and \citet{mandal2026}.
Their mode amplitudes and dynamical properties are extracted from the simulations and compared with the solar observations.

The organization of the paper is as follows. 
In Sect.~\ref{sec:methods}, the simulation of \citetalias{hotta2025} is briefly overviewed.
The results are presented in Sect.~\ref{sec:results}.
We discuss the implications in Sect.~\ref{sec:discussion} and the conclusions are summarized in Sect.~\ref{sec:summary}.


\section{Methods} \label{sec:methods}

\subsection{Simulation data: Overview of Hotta (2025)} \label{method:H25}

\begin{figure*}
\begin{center}
\includegraphics[width=0.75\linewidth]{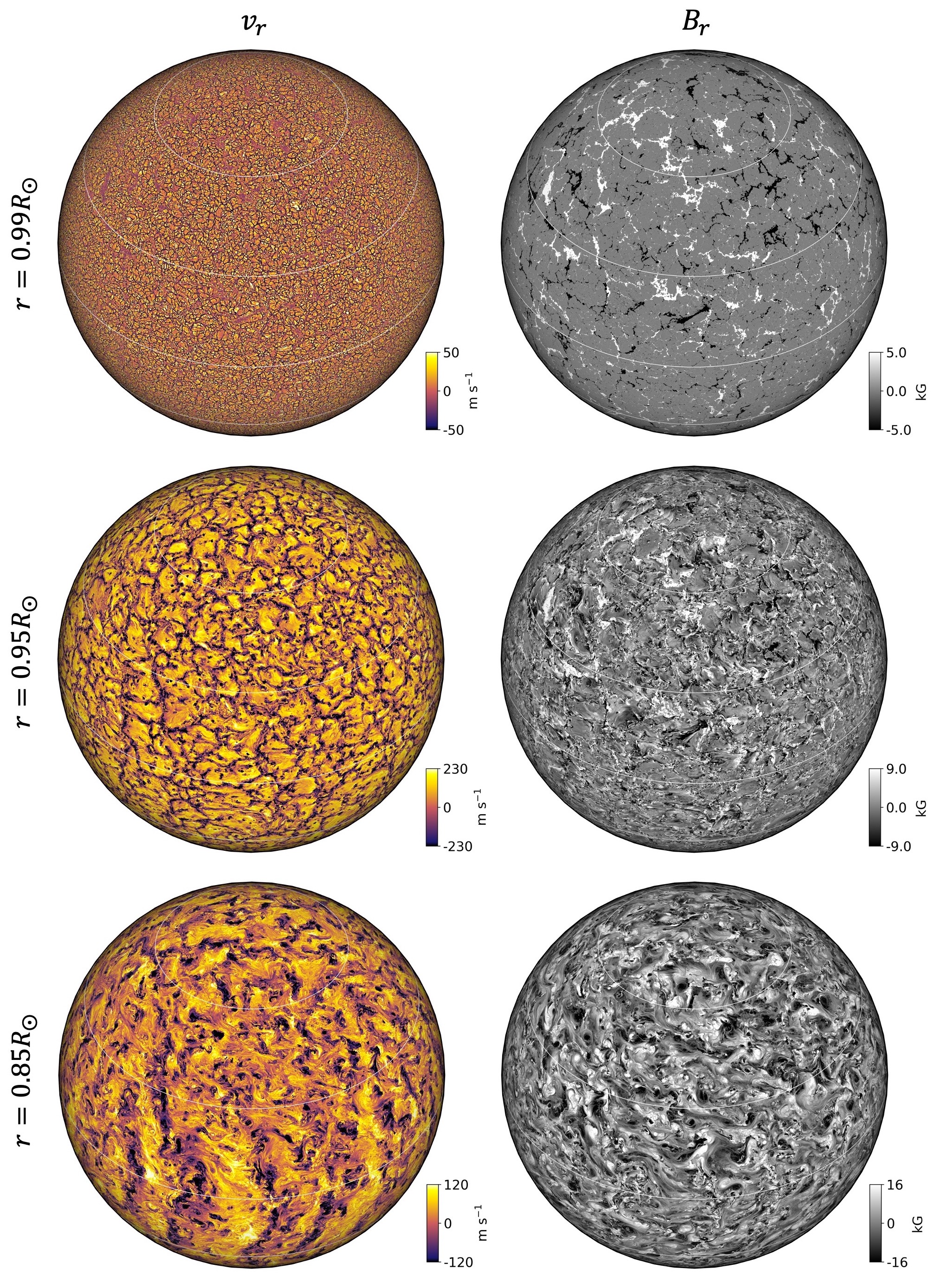}
\caption{
Snapshots of (\textit{left}) radial velocity $v_r$ and (\textit{right}) radial magnetic field $B_r$ from the \citetalias{hotta2025} simulation in the statistically stationary state ($t=13.7$~yr).
Top, middle, and bottom panels show the radial slices at the surface ($r=0.99R_{\odot}$), subsurface ($r=0.95R_{\odot}$), and in the middle convection zone ($r=0.85R_{\odot}$), respectively. 
}
\label{fig:snapshots}
\end{center}
\end{figure*}
\begin{figure*}
\begin{center}
\includegraphics[width=0.95\linewidth]{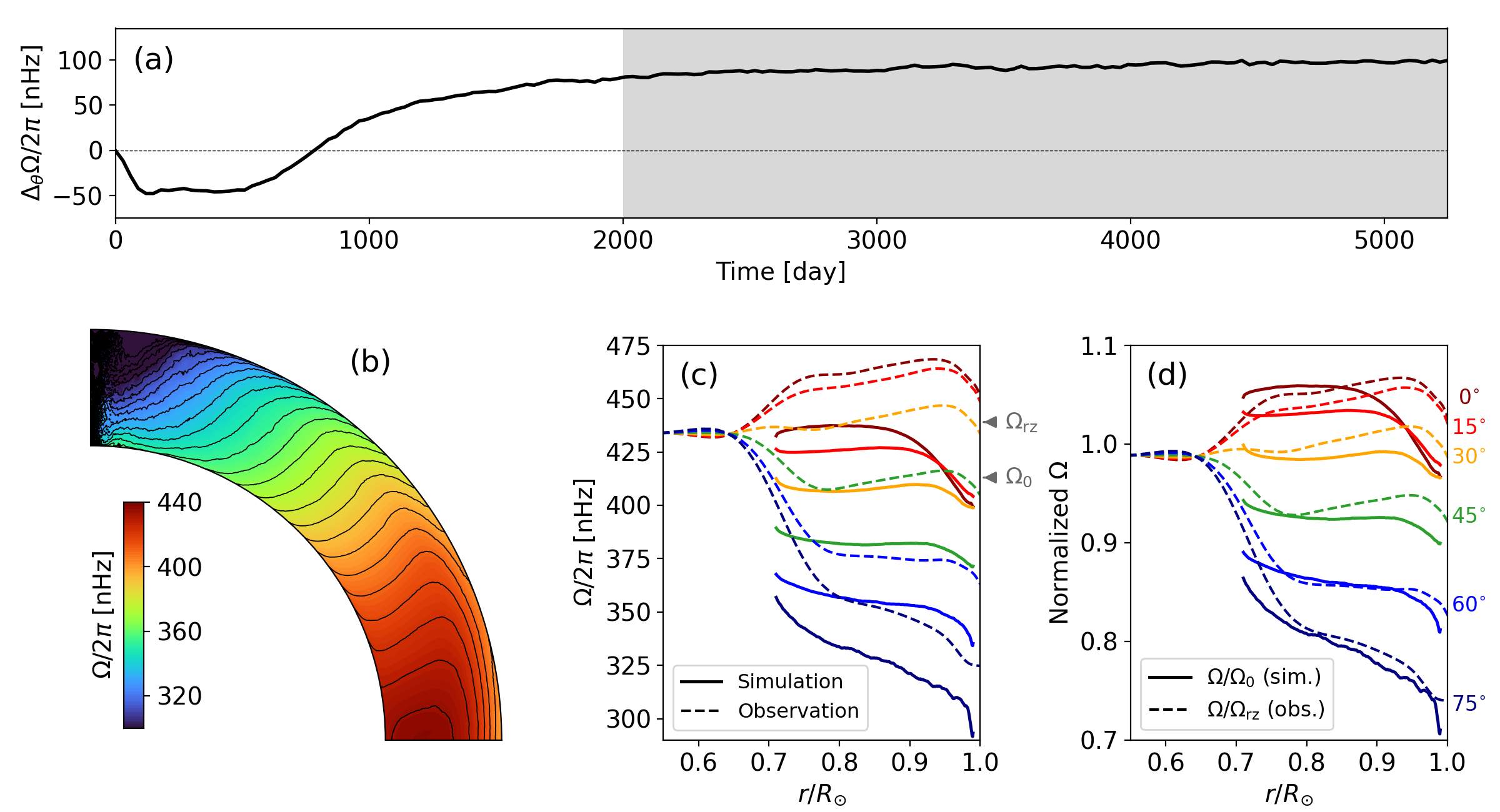}
\caption{
Differential rotation, $\Omega = \Omega_0 + \langle v_\phi \rangle /(r\sin{\theta})$, established in the \citetalias{hotta2025} simulation, where $v_\phi$ is the longitudinal velocity in the frame rotating at $\Omega_0$, and $\langle\cdot\rangle$ denotes the longitudinal average.
(a) Temporal evolution of the latitudinal differential rotation, $\Delta_\theta \Omega = \Omega_{\rm eq} - \Omega_{\rm pole}$, in the middle convection zone ($r=0.85R_{\odot}$).
Here, the rotation rates at the equator $\Omega_{\rm eq}$ and near the poles $\Omega_{\rm pole}$ are evaluated at $0^\circ$ and $70^\circ$ in latitude, respectively.
Positive (negative) $\Delta_\theta \Omega$ represents solar-like (anti-solar) differential rotation with faster (slower) equator.
The gray shaded area denotes the time period analyzed in this study.
(b) Two-dimensional (2D) profile of the differential rotation in a statistically stationary state.
(c) Radial profiles of differential rotation at selected latitudes.
Solid and dashed lines denote the \citetalias{hotta2025} simulation and the helioseismic observation \citep[from][]{larson2018}.
(d) Same as panel c, but with the rotation rates normalized by $\Omega_0$ for the simulation and by $\Omega_\mathrm{rz}$ for the observations, where $\Omega_0/2\pi = 413$~nHz and $\Omega_{\rm rz}/2\pi = 439$~nHz.
}
\label{fig:diffrot}
\end{center}
\end{figure*}
\begin{figure*}
\begin{center}
\includegraphics[width=0.9\linewidth]{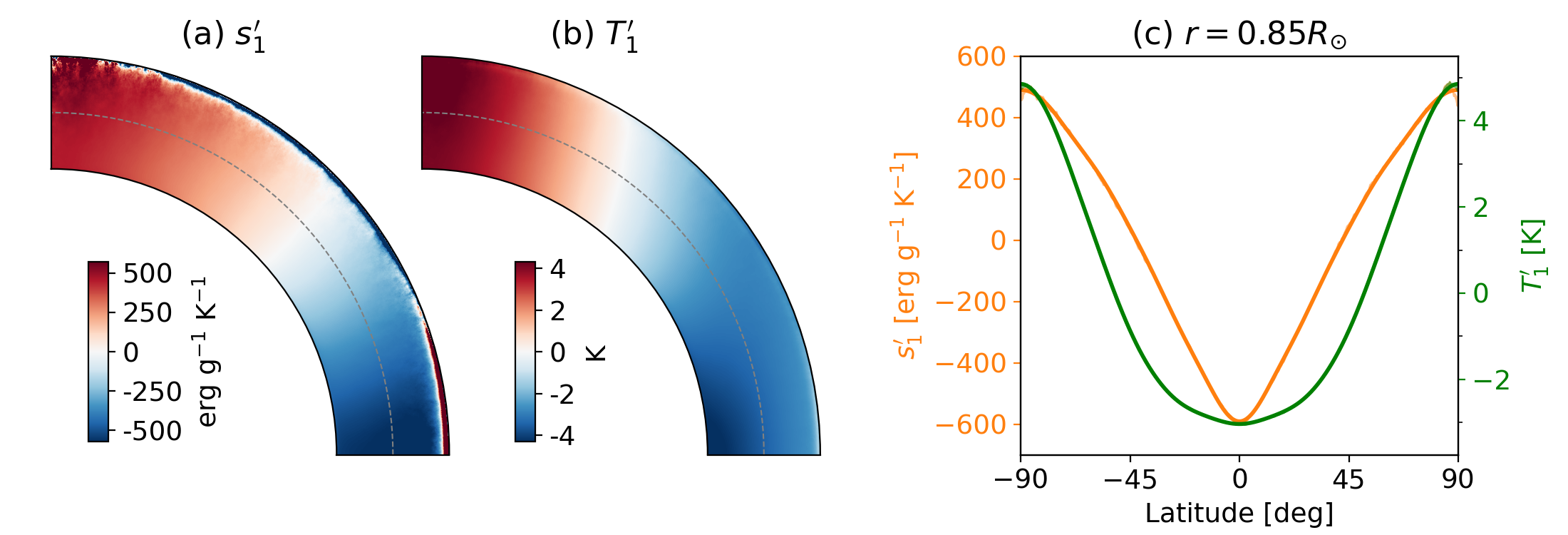}
\caption{
Profiles of entropy and temperature perturbations in the \citetalias{hotta2025} simulation.
(a, b) 2D profiles of entropy perturbation $s'_1$ and temperature perturbation $T'_1$ averaged over time and longitude, with spherically symmetric mean subtracted at each radius.
(c) Latitudinal variations of $s'_1$ (orange) and $T'_1$ (green) at the middle convection zone ($r=0.85R_{\odot}$).
}
\label{fig:s1T1}
\end{center}
\end{figure*}

In this paper, we analyze the unprecedentedly high-resolution simulation data published in \citetalias{hotta2025}.
By solving fully compressible magnetohydrodynamic (MHD) equations with the \texttt{R2D2} code \citep[see][for further details]{hotta2021}, the simulation has successfully reproduced the solar-like differential rotation with (almost) the solar rotation rate, $\Omega_0=2.595$~$\mu$rad\,s$^{-1}$, and the solar luminosity, $L_{\odot}=3.846\times 10^{33}$~erg~s$^{-1}$, with the help of vigorous turbulent dynamo action.
The numerical domain is a full spherical shell extending from the base of the convection zone ($r_{\rm min}=0.71R_{\odot}$) to $7$~Mm below the photosphere ($r_{\rm max}=0.99R_{\odot}$), where the background is stratified according to the standard model S \citep{christensen1996}.
In this simulation, no explicit diffusivities are used.
Instead, a slope-limited artificial diffusion scheme \citep[proposed by][]{rempel2014} is adopted, so that the effective viscous, thermal, and magnetic diffusivities depend on numerical resolution.
The simulation was performed on a Yin-Yang grid \citep{kageyama2004}, and the resulting data were remapped onto a conventional spherical grid with $(N_r, N_\theta, N_\phi)=(512,\ 4096,\ 8192)$ for the analysis in this study.
The effective Reynolds number of the simulation is of order $10^4$.
As illustrated in Fig.~\ref{fig:snapshots} showing representative snapshots of the radial velocity $v_r$ and radial magnetic field $B_r$, both the flows and magnetic fields exhibit highly turbulent structures spanning a broad range of spatial scales.

Starting from the zero initial condition, the simulation was run for a total of $5276$~days.
The solar-like differential rotation with faster equator and slower poles was established and became almost stationary from time $t \sim 2000$~days, although some temporal variations remained (Fig.~\ref{fig:diffrot}a).
The simulated differential rotation successfully reproduces several key observational features, including not only equatorial acceleration but also a significant departure from the Taylor–Proudman state and the presence of an NSSL  at all latitudes (Fig.~\ref{fig:diffrot}b).
However, because the simulation domain excludes the underlying radiative interior, the tachocline is not reproduced.
We also note that, due to the slightly lower system rotation rate adopted in the simulation ($\Omega_0/2\pi=413$~nHz), the overall rotation is marginally slower than in the real Sun by $\sim 25$~nHz (Fig.~\ref{fig:diffrot}c).
Figure~\ref{fig:diffrot}d shows that the scaled differential rotation, in which this effect is accounted for, is in excellent agreement with the observations.

The analyses by \citetalias{hotta2025} have revealed that the bulk of the convection zone ($r \lesssim 0.9R_{\odot}$) is in thermal wind balance, where the Coriolis torque associated with the non–Taylor–Proudman differential rotation is balanced by the baroclinic torque arising from the latitudinal entropy gradient \citep[e.g.,][]{miesch2006}.
In contrast, the thermal wind balance breaks down in the NSSL, where Reynolds and Maxwell stresses play a dominant role.
As shown in Fig.~\ref{fig:s1T1}a, the latitudinal entropy gradient is negative in the northern hemisphere throughout most of the convection zone.
Within a very thin layer near the surface, however, the entropy perturbation exhibits a distinct behavior, which is likely attributable to surface magnetic fields (Bekki \& Hotta, in prep.).
The pole-to-equator temperature difference corresponding to the latitudinal entropy variation in the bulk convection zone is approximately $-7.5$~K (Fig.~\ref{fig:s1T1}b and c), which is expected to be sufficiently large to induce baroclinic instability \citep[][]{bekki2022a,bekki2024b}.

In addition to low-cadence data saved at $30$-day intervals analyzed in \citetalias{hotta2025}, high-cadence data with a temporal resolution of $\Delta t = 1$~day are also available.
In this dataset, spherical shell slices of each variable are available at six radial points, i.e.,  $r=0.71R_{\odot},\ 0.8R_{\odot}, \ 0.85R_{\odot}, \ 0.9R_{\odot}, \ 0.95R_{\odot},$ and $0.99R_{\odot}$.
In this study, we mainly analyze this high-cadence shell slice data over the later phase of the simulation, spanning the range $2000-5276$~days ($\sim 9$~years), during which the solar-like differential rotation is almost statistically stationary (the shaded period in Fig.~\ref{fig:diffrot}a).
The total number of temporal data points is thus $N_t=3277$.

\begin{figure*}
\begin{center}
\includegraphics[width=0.75\linewidth]{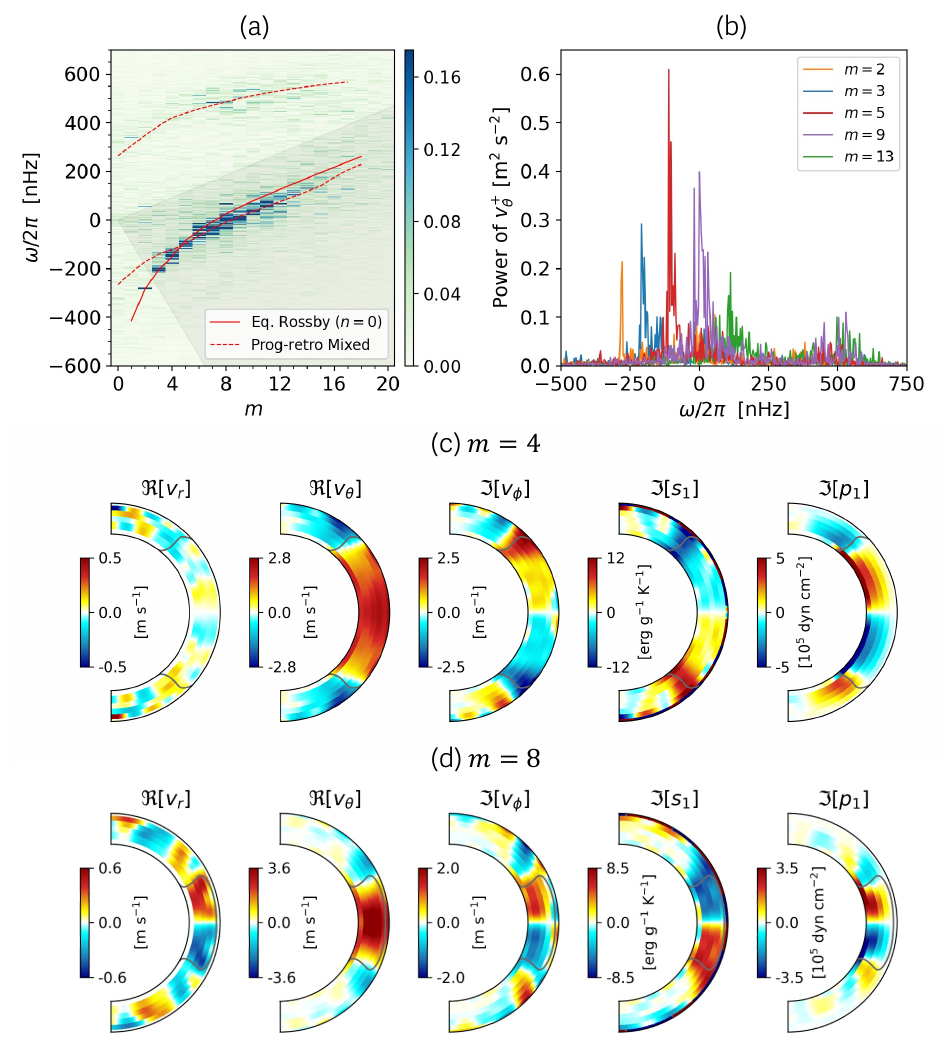}
\caption{
Equatorial Rossby modes in the \citetalias{hotta2025} simulation.
(a) Equatorial power spectrum (averaged over $\pm 15^{\circ}$ in latitude) of latitudinal velocity $|\tilde{v}^{+}_{\theta}|^2$ in the middle convection zone (averaged between $0.8R_{\odot} \leq r \leq 0.9R_{\odot}$).
Red solid and dashed curves show the linear dispersion relations of the $n=0$ equatorial Rossby modes and the mixed prograde-retrograde inertial modes computed from the linear model (see Sect.~\ref{sec:linear}).
All the frequencies are measured in a frame rotating at $\Omega_0/2\pi = 413$~nHz.
The gray shaded area show the frequency range in which the modes can be significantly affected by the differential rotation via corotation resonance, $\Omega_{\rm pole} \leq \omega / m +\Omega_0 \leq \Omega_{\rm eq}$, where $\Omega_{\rm eq}/2\pi = 436$~nHz and $\Omega_{\rm pole}/2\pi = 320$~nHz.
(b) Same power spectra at selected azimuthal orders $m$.
(c,d) Extracted eigenfunctions of the equatorial Rossby modes at $m=4$ and $m=8$, respectively.
The longitudes corresponding to the real and imaginary phases are $\phi=\phi_0$ and $\phi=\phi_0-\pi/2m$, respectively, where $\phi_0$ denotes the longitude at which $v_\theta$ reaches its maximum at the equator.
Gray solid lines denote the locations of the viscous critical layers where the mode phase speed matches the local differential rotation speed, i.e., $\omega - m[\Omega(r,\theta)-\Omega_0]=0$ (corotation resonances).
}
\label{fig:eig_eqRos}
\end{center}
\end{figure*}
\begin{figure}
\begin{center}
\includegraphics[width=0.95\linewidth]{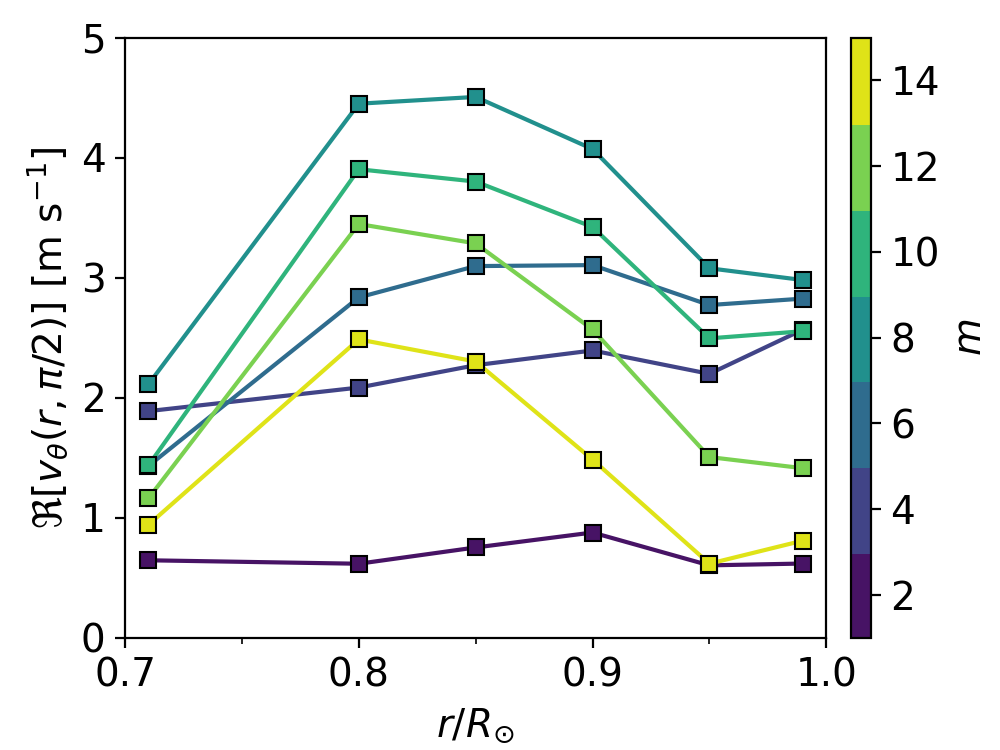}
\caption{
Radial structures of the latitudinal velocity eigenfunctions of the extracted equatorial Rossby modes at the equator.
Different colors correspond to different azimuthal orders $m$.
Only real parts of the eigenfunctions, $\Re[v_\theta]$, are shown, where the real phase is defined such that $v_\theta$ becomes maximum.
}
\label{fig:eig_eqRos_rad}
\end{center}
\end{figure}
\begin{figure*}
\begin{center}
\includegraphics[width=0.85\linewidth]{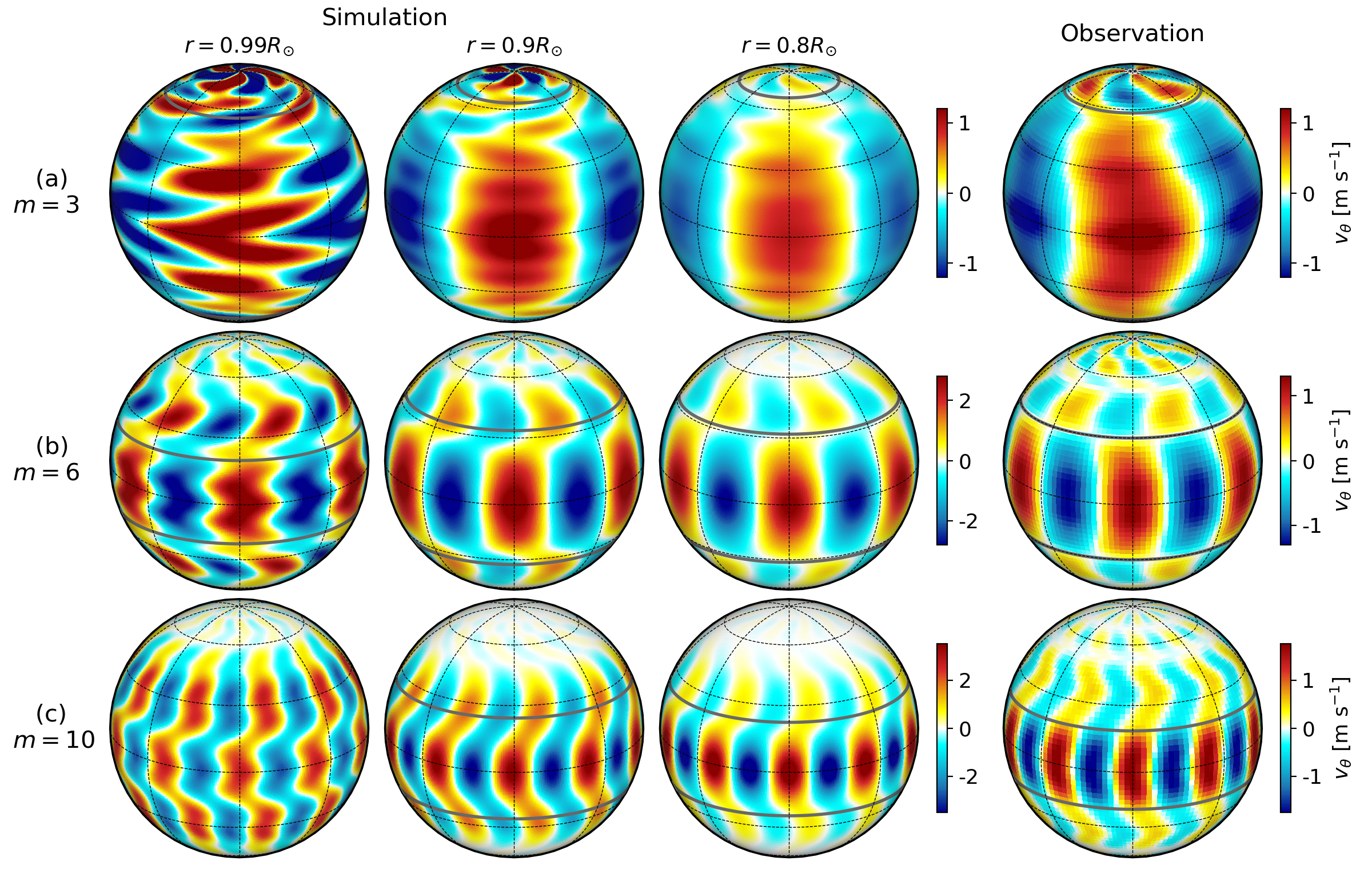}
\caption{
Height dependence of the latitudinal velocity ($v_\theta$) eigenfunctions of the equatorial Rossby modes with (a) $m=3$, (b) $m=6$, and (c) $m=10$.
The first three columns show the eigenfunctions extracted from the \citetalias{hotta2025} simulation at $r=0.99R_\odot$, $0.9R_\odot$, and $0.8R_\odot$, respectively. 
The rightmost column shows the observed surface eigenfunctions from \citet{mandowara2026}.
Gray solid lines show the latitudes of the critical layers.
We note that, in the \citetalias{hotta2025} simulation, critical layers are absent at $r=0.99R_\odot$ for the equatorial Rossby modes with $m \geq 7$.
}
\label{fig:eig_eqRos_surf}
\end{center}
\end{figure*}

\begin{figure*}
\begin{center}
\includegraphics[width=0.75\linewidth]{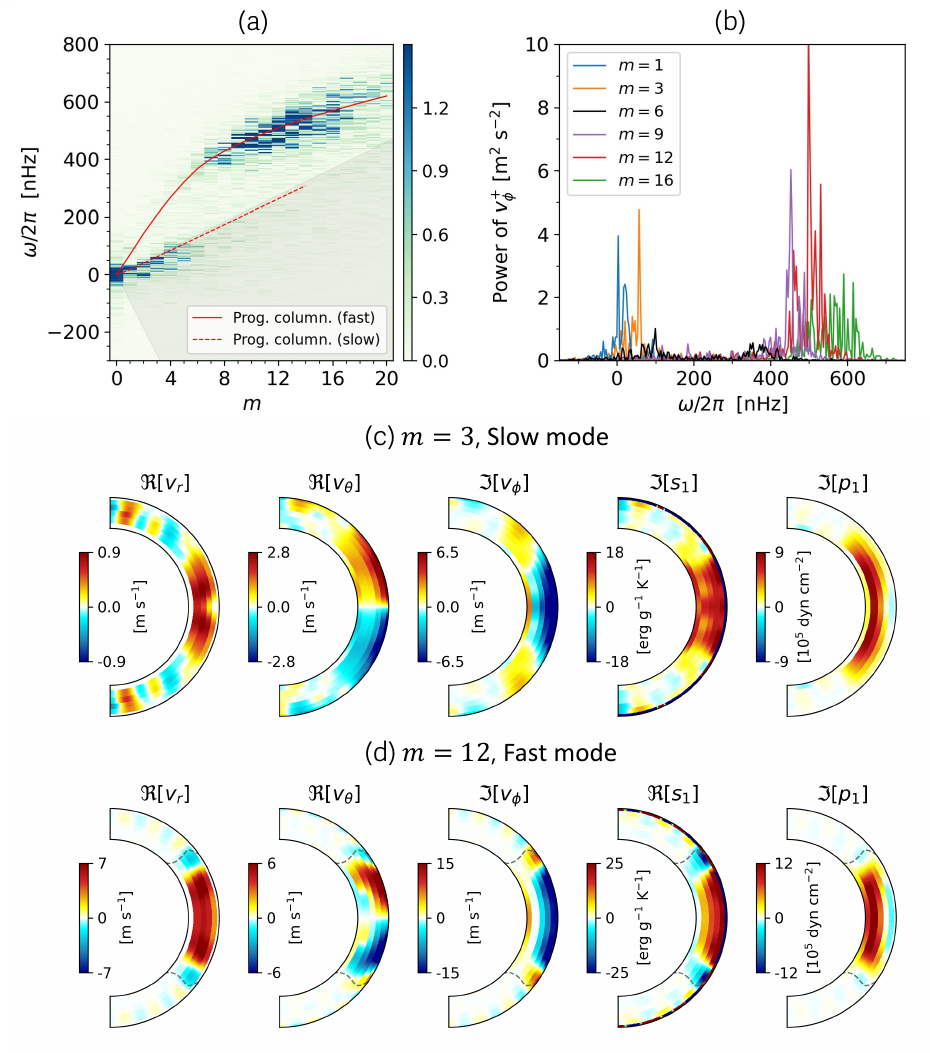}
\caption{
Prograde columnar modes in the \citetalias{hotta2025} simulation.
(a) Equatorial power spectrum of longitudinal velocity $|\tilde{v}^{+}_{\phi}|^2$ in the middle convection zone (averaged between $0.8R_{\odot} \leq r \leq 0.9R_{\odot}$).
Red solid and dashed curves show the linear dispersion relations of the fast and slow prograde columnar modes computed from the linear model.
All the frequencies are measured in a frame rotating at $\Omega_0/2\pi = 413$~nHz.
(b) Same power spectra at selected azimuthal orders $m$.
(c) Extracted eigenfunctions of the slow mode at $m=3$.
Note that the imaginary part of the entropy eigenfunction, $\Im[s_1]$, is shown to highlight the quadrature phase relation between $v_r$ and $s_1$.
(d) The same as panel c but for the fast mode at $m=12$.
Note that the real part of the entropy eigenfunction, $\Re[s_1]$, is shown to highlight its convective nature.
Gray dashed curves denote the locations of the turning surfaces, i.e., $|\omega - m(\Omega-\Omega_0)| \lesssim 2\Omega$ (under weak differential rotation approximation).
}
\label{fig:eig_thRos}
\end{center}
\end{figure*}
\begin{figure*}
\begin{center}
\includegraphics[width=0.85\linewidth]{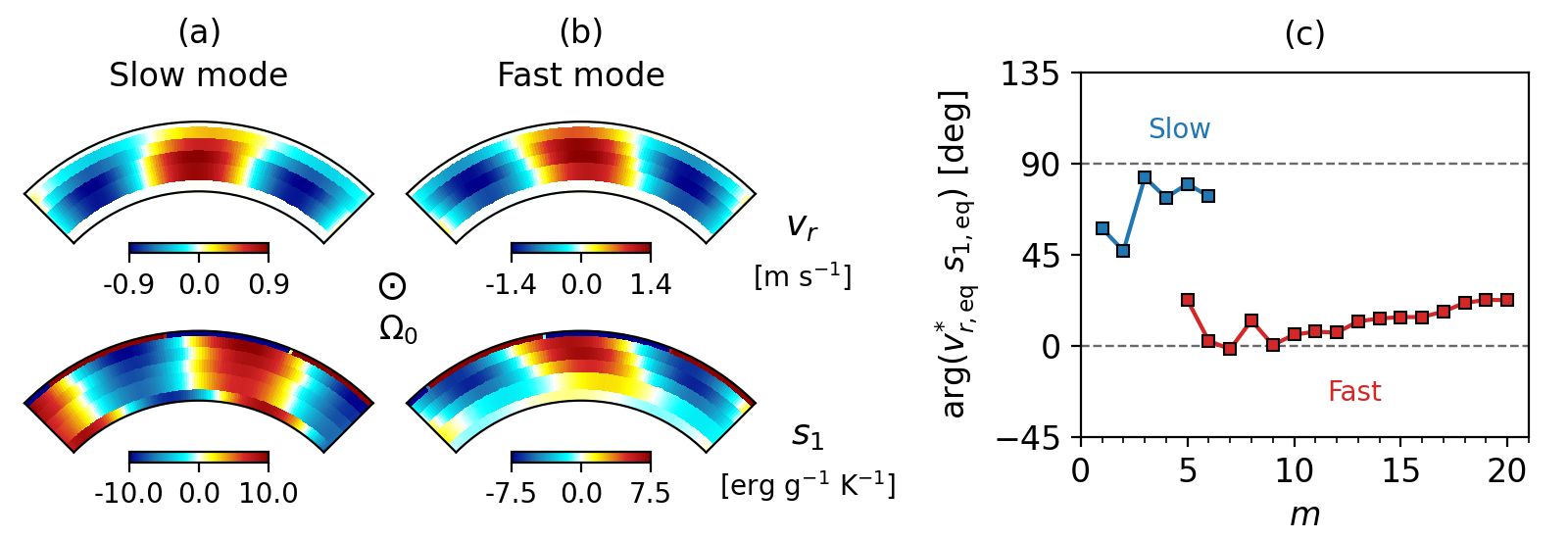}
\caption{
Phase differences between radial velocity $v_r$ and entropy perturbation $s_1$ in the low-frequency columnar modes (at $m \lesssim 6$) and the prograde-propagating columnar modes (at $m \gtrsim 6$).
(a) Extracted eigenfunctions of $v_r$ and $s_1$ of the $m=6$ low-frequency columnar mode in the equatorial plane seen from the north pole.
(b) Same as panel a but for the $m=6$ prograde columnar mode.
(c) Phase difference, $\mathrm{arg}(v_{r, \rm eq}^{*} s_{1, \rm eq})$, of the extracted eigenfunctions plotted against $m$. Here, $v_{r,\rm eq}$ and $s_{1,\rm eq}$ are the eigenfunctions at the equator averaged over height.
}
\label{fig:thRos_phase}
\end{center}
\end{figure*}
\begin{figure*}
\begin{center}
\includegraphics[width=0.75\linewidth]{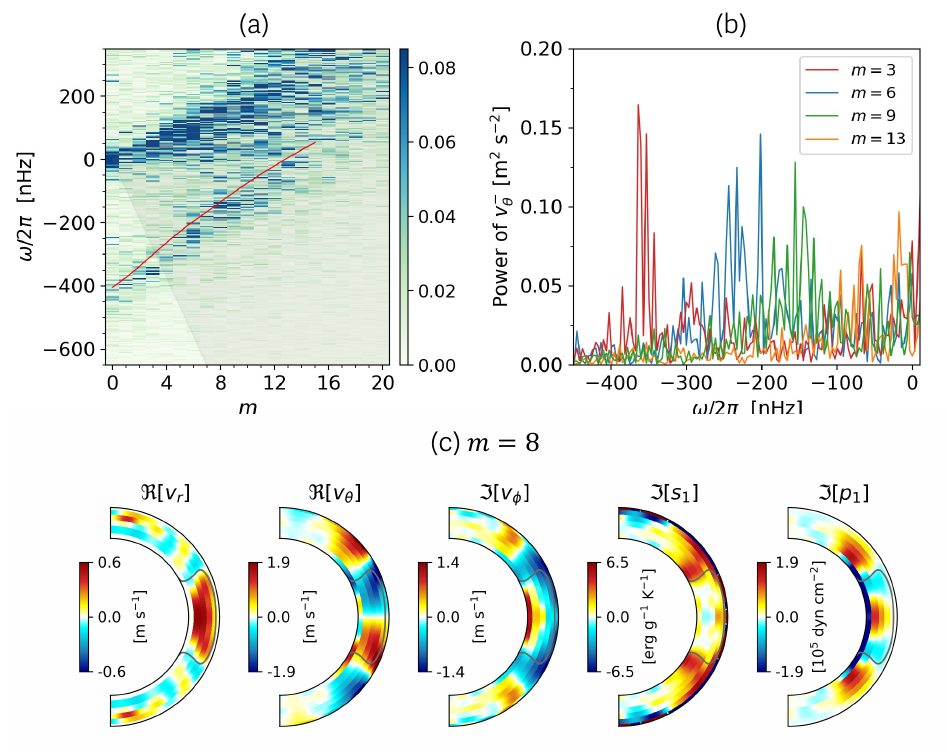}
\caption{
High-frequency retrograde (HFR) modes identified in the \citetalias{hotta2025} simulation.
(a) Equatorial power spectrum of the north-south anti-symmetric component of the latitudinal velocity $|\tilde{v}^{-}_{\theta}|^2$ in the middle convection zone.
Red solid curve show the linear dispersion relation of the HFR modes obtained from the linear model.
All the frequencies are measured in a frame rotating at $\Omega_0/2\pi = 413$~nHz.
(b) Same power spectra at selected azimuthal orders $m$.
(c) Extracted eigenfunctions of the HFR mode at $m=8$.
Gray solid lines denote the locations of the viscous critical layers where the corotation resonances occur.
}
\label{fig:eig_HFR}
\end{center}
\end{figure*}

\begin{figure*}
\begin{center}
\includegraphics[width=0.75\linewidth]{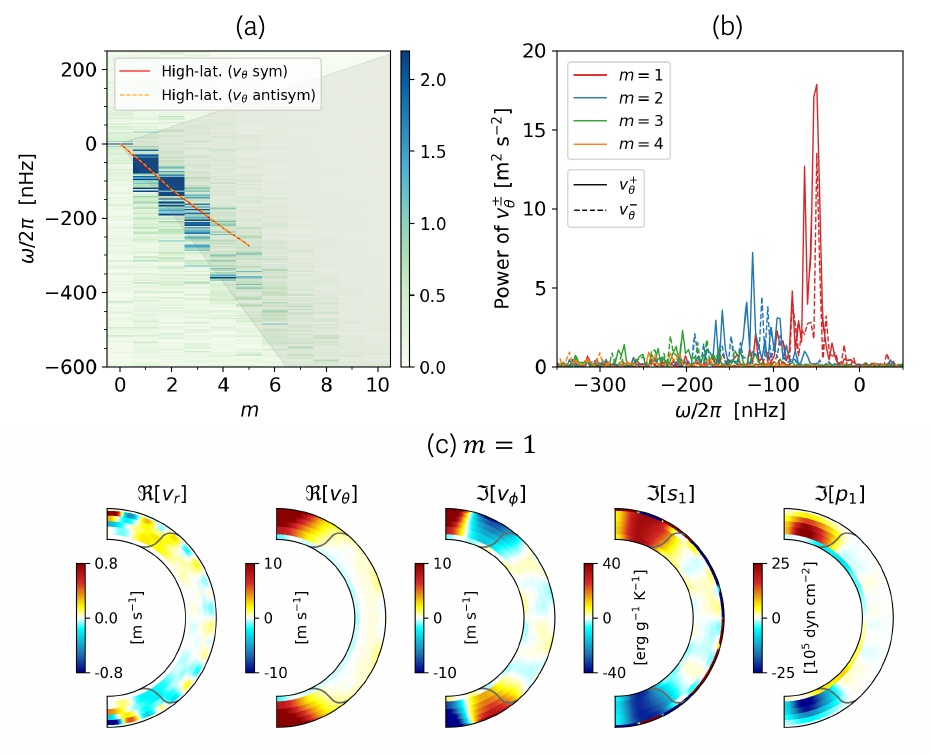}
\caption{
High-latitude modes identified in the \citetalias{hotta2025} simulation.
(a) High-latitude power spectrum (averaged between $75^{\circ}-90^{\circ}$ in latitude) of the latitudinal velocity $v_{\theta}$ in the middle convection zone.
Red solid and orange dashed curves show the linear dispersion relations of the high-latitude modes with north-south symmetric and antisymmetric $v_{\theta}$, respectively.
All the frequencies are measured in a frame rotating at $\Omega_0/2\pi = 413$~nHz.
(b) High-latitude power spectra of $v_{\theta}^{+}$ (solid) and $v_{\theta}^{-}$ (dashed) at selected azimuthal orders $m$.
(c) Extracted eigenfunctions of the $m=1$ mode with north-south symmetric $v_{\theta}$ and antisymmetric $v_{\phi}$.
}
\label{fig:eig_HL}
\end{center}
\end{figure*}

\begin{figure}
\begin{center}
\includegraphics[width=0.93\linewidth]{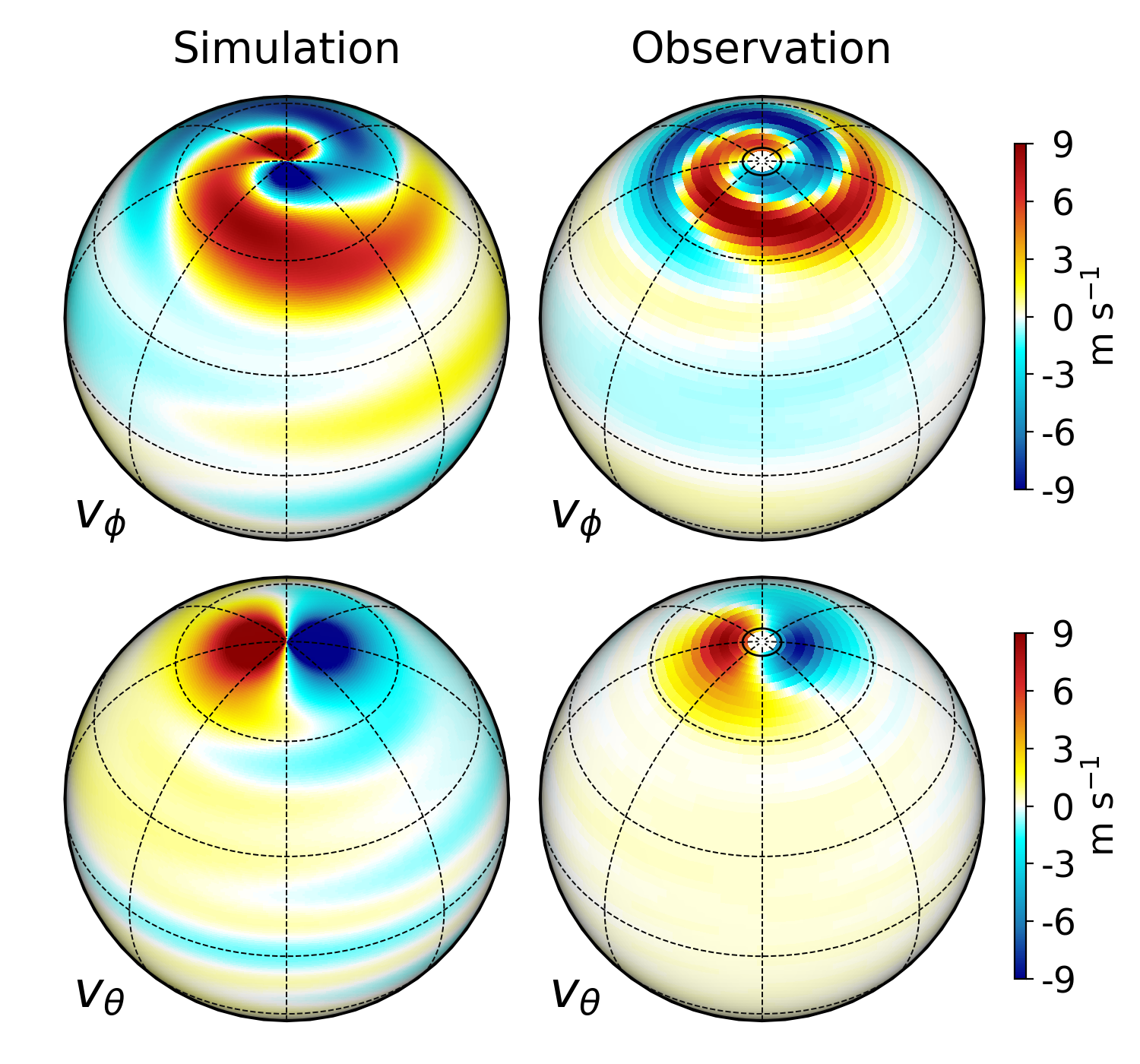}
\caption{
Extracted eigenfunction near the surface ($r=0.99R_{\odot}$) of the $m=1$ high-latitude mode with north-south symmetric $v_{\theta}$, compared with the observation \citep[right panels, from][]{mandowara2026}.
Top and bottom rows show the surface eigenfunctions of the longitudinal velocity $v_\phi$ and the latitudinal velocity $v_\theta$, respectively.
}
\label{fig:spiral_m1HL}
\end{center}
\end{figure}

\subsection{Power spectra} \label{method:power}

To study large-scale inertial modes in the low-frequency range, we first apply Fourier transforms in time and longitude
\begin{eqnarray}
    && \tilde{q}(r,\theta;m,\omega) = \frac{1}{N_t N_\phi} \sum_{t,\phi} q(r,\theta,\phi,t) e^{\ii (\omega t - m\phi)},
\end{eqnarray}
where $q$ denotes an arbitrary variable to be analyzed (e.g., the velocity components $v_{r}$, $v_\theta$, and $v_\phi$; the perturbations of entropy $s_1$, gas pressure $p_1$, or magnetic pressure $p_{\rm mag}$), $m$ is the azimuthal order, and $\omega$ is the angular frequency measured in a frame corotating at $\Omega_0$.

For each inertial-mode family, we first choose a representative physical variable and average its Fourier power over the latitude range where the target mode is most clearly visible (cf. Figs.~\ref{fig:eig_eqRos}a, \ref{fig:eig_thRos}a, \ref{fig:eig_HFR}a, and \ref{fig:eig_HL}b).
To measure the mode frequencies and linewidths, we perform a Lorentzian fit to the latitudinally averaged power spectra in the middle convection zone at fixed $m$.
We use the functional form
\begin{eqnarray}
    && P_{\rm fit}(\omega) = \frac{A}{1+[(\omega-\omega_0)/(\Gamma/2)]^2} + B, \label{eq:lorentzian}
\end{eqnarray}
where $A$ is the height of the Lorentzian, $\omega_0$ is the mode frequency, $\Gamma$ is the full width at half maximum, and $B$ is the background power.
The fitting was performed using the maximum likelihood method under the assumption that each point in the power spectrum is drawn from an exponential probability density function \citep[][]{anderson1990,lekshmi2026}.

\subsection{Eigenmode extraction: Singular-value-decomposition} \label{method:svd}

Once the inertial modes have been identified in the power spectra, their spatial eigenfunctions are extracted using a singular-value decomposition (SVD) following the method of \citet[][]{bekki2022b,bekki2024b}.
The complex Fourier coefficients of a representative variable are decomposed by SVD within a selected radial and frequency range around the target mode.
For each fixed $m$, we retain the first right singular vector, $V_0(\omega)$, associated with the first (most dominant) singular value, and extract the spatial eigenfunctions of all physical variables by projecting their complex Fourier coefficients onto $V_0(\omega)$ at all depths.
Further details of the procedure are given in \citet[][in Sect.~3.2]{bekki2022b} and \citet[][in Supplementary Materials]{bekki2024b}.
Because the extracted eigenfunctions generally contain small-scale fluctuations associated with highly turbulent motions in the \citetalias{hotta2025} simulation, we smooth them by truncating their Legendre  expansions at $\ell = 20$.

\subsection{Linear eigenmode analysis} \label{sec:linear}

For mode identification, we make use of the linear model of solar inertial modes developed by \citet{bekki2022a}. 
The linear eigenmode solver adopts the same computational domain, boundary conditions, and background stratification as the \citetalias{hotta2025} simulation.
The simulated differential rotation (Fig.~\ref{fig:diffrot}b) and latitudinal entropy gradient (Fig.~\ref{fig:s1T1}a) are also included into the model after being smoothed by truncating their Legendre expansions at $\ell = 20$ for numerical stability.
For simplicity of the analysis, effects of the meridional circulation are ignored \citep[][]{fournier2022} and the background entropy stratification in the radial direction is assumed to be adiabatic ($\delta=0$) throughout the convection zone (see Appendix~\ref{appendix:delta}).
The turbulent viscous and thermal diffusivities are estimated using the first-order smoothing approximation \citep[FOSA, e.g.,][]{moffatt1978} as
\begin{eqnarray}
&& \nu_{\rm t}=\kappa_{\rm t} =\frac{\tau_{\rm c}}{3}v_{\rm rms}^2 =\frac{v_{\rm rms}H_{\rm p}}{3}, \label{eq:fosa}
\end{eqnarray}
where $v_{\rm rms}$ is the root-mean-square (rms) convective velocity in the \citetalias{hotta2025} simulation, $H_{\rm p}$ is the pressure scale height, and $\tau_{\rm c}=H_{\rm p}/v_{\rm rms}$ is the convective turnover timescale.
For reference, the resulting turbulent diffusivities become $\sim 8.9\times 10^{12}$~cm$^2$~s$^{-1}$, when averaged over the whole convection zone.

\section{Results} \label{sec:results}

\begin{figure*}
\begin{center}
\includegraphics[width=0.95\linewidth]{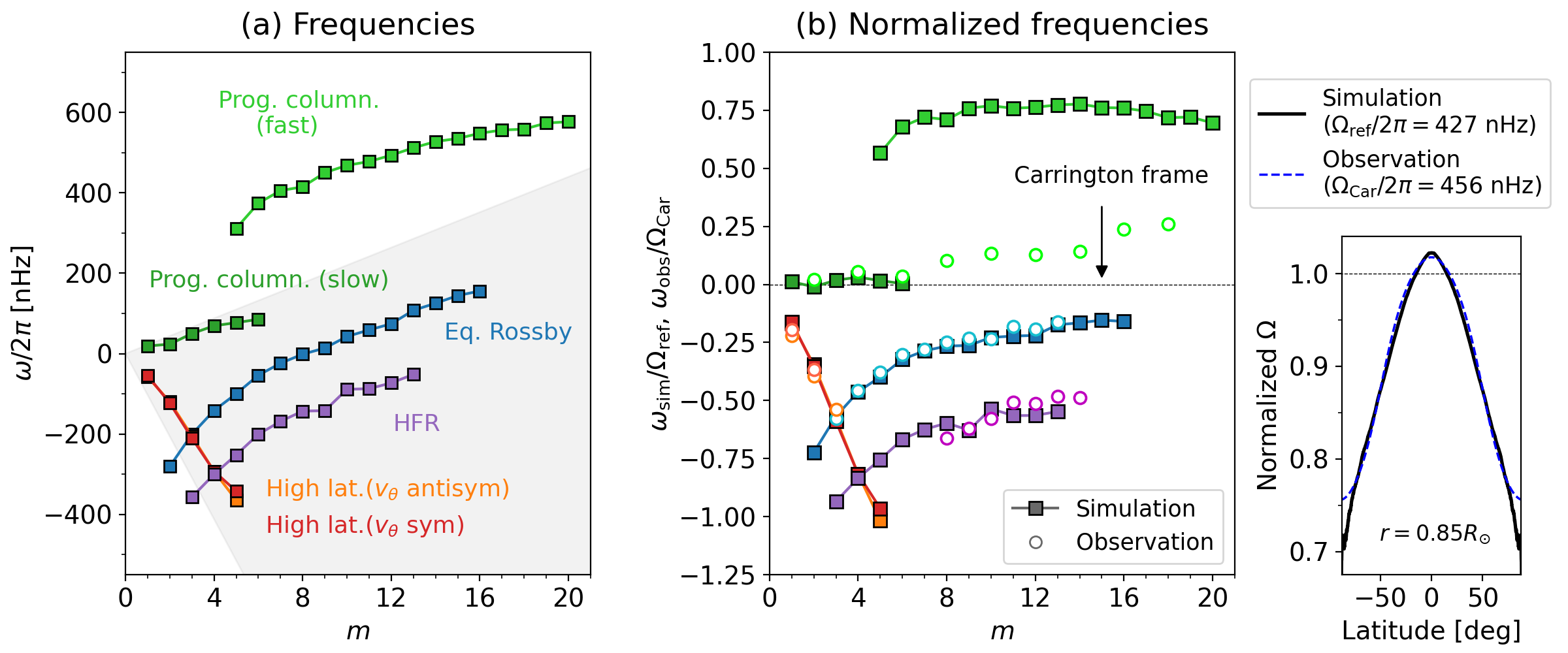}
\caption{
Frequencies of the inertial modes identified in the \citetalias{hotta2025} simulation.
(a) Mode frequencies measured in a frame corotating at $\Omega_0/2\pi = 413$~nHz.
Different colors represent different types of inertial modes.
The gray shaded area show the frequency range in which the modes can be significantly affected by the differential rotation via corotation resonance, $\Omega_{\rm pole} \leq \omega / m +\Omega_0 \leq \Omega_{\rm eq}$, where $\Omega_{\rm eq}/2\pi = 436$~nHz and $\Omega_{\rm pole}/2\pi = 320$~nHz.
(b) Comparison of the normalized frequencies between the simulation, $\omega_{\rm sim} / \Omega_{\rm ref}$, (square symbols) and the observations, $\omega_{\rm obs} / \Omega_{\rm Car}$, (open circles).
Here, we use the frame corotation rates $\Omega_{\rm ref}/2\pi=427$~nHz for the simulation, and $\Omega_{\rm Car}/2\pi=456$~nHz (Carrington rotation rate) for the observation.
When such normalized, the simulated and the observed differential rotation profiles in the middle convection zone $\Omega(0.85R_{\odot},\theta)/\Omega_{\rm ref}$ show an excellent agreement, as shown in the rightmost panel.
The observed frequencies are taken from \citet[][for equatorial Rossby and high-latitude modes]{mandowara2026}, from \citet[][for HFR modes]{hanson2022}, and from \citet[][for thermal Rossby modes]{mandal2026}. The latter two have been corrected to the Carrington frame.
}
\label{fig:disp_comp}
\end{center}
\end{figure*}

\subsection{Equatorial Rossby modes} \label{results:eqRos}

Figure~\ref{fig:eig_eqRos}a shows the $m$-$\omega$ power spectrum of the north-south symmetric component of the latitudinal velocity $v_{\theta}^{+}$ averaged near the equator ($\pm 15\degree$ in latitude) in the middle convection zone.
A ridge of excess power corresponding to the equatorial Rossby modes can be seen for the range of azimuthal orders, $2 \leq m \lesssim 15$.
The power is broadly distributed in frequency space, roughly consistent with the linear dispersion relations of both the equatorial Rossby modes with no radial node ($n=0$) and those with one radial node ($n=1$), the latter of which are understood as retrograde branch of the mixed prograde-retrograde inertial modes \citep[][]{bekki2022a,jain2024}.
Figure~\ref{fig:eig_eqRos}b shows the power spectrum at five selected values of the azimuthal order $m$.
Sharp prominent peaks corresponding to the equatorial Rossby modes are evident in all power spectra.
A weaker excess of power is also seen at higher  frequencies  $\omega/2\pi \sim 400-600$~nHz, corresponding to the prograde branch of the mixed inertial modes \citep[][]{bekki2022b}, which we do not discuss  further in this study.

We then extracted the spatial eigenfunctions of the equatorial Rossby modes using the method described in Sect.~\ref{method:svd}. 
Meridional eigenfunctions (cross sections at selected longitudinal phase) for the representative $m=4$ and $m=8$ modes are shown in Figs.~\ref{fig:eig_eqRos}c and d. 
As theory predicts \citep[][]{bekki2022a}, these modes are quasi-toroidal at relatively low $m$, with radial motions significantly weaker than the horizontal motions.
It is also seen that the latitudinal velocity eigenfunction changes its sign near the latitudes of the  critical layers (indicated by the gray curves), where corotation resonances occur, i.e. where $\omega=m[\Omega(r,\theta)-\Omega_0]$ \citep[][]{gizon2020}. 
The  characteristic behavior has been reported in the observations \citep[][]{ loeptien2018,proxauf2020,mandowara2026}. 
We find that, as $m$ increases, the equatorial Rossby modes involve non-negligible radial motions near the latitudes of the critical layers, with amplitudes reaching $\sim20-30\%$ of the maximum horizontal motions, and are therefore no longer toroidal.

Figure~\ref{fig:eig_eqRos_rad} shows the radial structure of the latitudinal velocity eigenfunctions at the equator, $\Re[v_{\theta}(r,\pi/2)]$, for different $m$. 
It clearly indicates that the equatorial Rossby modes detected in the \citetalias{hotta2025} simulation correspond primarily to the $n=0$ modes, as evidenced by the absence of radial nodes.
For $m \gtrsim 6$, their radial eigenfunctions tend to peak in the middle of the convection zone, in contrast to the classical $r^m$ dependence expected for inviscid sectoral Rossby modes under uniform rotation \citep[][]{provost1981,damiani2020}.
This deviation may be attributed to the effects of differential rotation, which induces non-toroidal motions near the critical layers, and relatively strong turbulent diffusion in the \citetalias{hotta2025} simulation, which can also modify their radial eigenfunctions \citep[][]{bekki2022a,bekki2022b}.

Figure~\ref{fig:eig_eqRos_surf} shows the spatial patterns of the $v_\theta$ eigenfunctions on spherical shells. 
Near the upper boundary, the extracted eigenfunctions exhibit complex spatial structure, possibly due to contamination by convective noise. 
At greater depths, the global-scale patterns become clearer and closely resemble the observed surface eigenfunctions, characterized by a nearly sectoral profile peaking at the equator at latitudes lower than those of the critical layers and the non-sectoral behaviors at higher latitudes.
Differences in the mode amplitudes between the simulation and the observations are discussed in Sect.~\ref{results:amp}.

\subsection{Prograde columnar modes: Fast and slow modes} \label{results:thRos}

Next, we discuss the prograde columnar modes, which appear as the dominant large-scale flow pattern in the simulation and are manifested as north-south aligned downflow lanes near the equator. 
Such prograde-propagating convective columns aligned with the rotation axis have been studied theoretically \citep[e.g.,][]{busse1970,busse2002} and repeatedly seen in numerical simulations \citep[e.g.,][]{miesch2000,miesch2008,hindman2020}.
In the Sun's convection zone, this prograde propagation is caused by the compressional $\beta$ effect arising from the background density stratification \citep[e.g.,][]{glatzmaier1981,verhoeven2014,bekki2025}. 

These modes are most clearly visible in the power spectrum of the equatorially symmetric longitudinal velocity $v_{\phi}^{+}$, as shown in Fig.~\ref{fig:eig_thRos}a. 
To aid the interpretation of the simulation, we overplot the dispersion relations of the two classes of prograde columnar modes obtained from the linear model: 
The first, shown by the red solid curve, corresponds to \textit{fast} modes propagating at relatively high prograde frequencies. They are the classical prograde columnar modes (also known as thermal Rossby modes), which originate from the compressional $\beta$ effect associated with the background density stratification and have been extensively discussed in the literature \citep[][]{glatzmaier1981,bekki2022a,hindman2022}. 
On the other hand, the second, shown by the red dashed curve, corresponds to \textit{slow} modes with much lower frequencies. Note that their frequencies become nearly zero when measured in the frame co-rotating at the equatorial rotation rate. 
These slow modes are intrinsically thermally diffusive in the linear model (see Appendix~\ref{appendix:slow} for more details).
The simulated power spectrum exhibits two distinct power ridges corresponding to the fast and slow modes. 
At low azimuthal orders ($m \lesssim 6$), most of the power is concentrated in the slow-mode branch, whereas the power is concentrated in the fast-mode branch at $m\gtrsim6$.

The same power spectra for selected values of $m$ are shown in Fig.~\ref{fig:eig_thRos}b, where sharp peaks corresponding to the individual prograde columnar modes are clearly identified.
Figures~\ref{fig:eig_thRos}c and d show representative eigenfunctions extracted for the $m=3$ slow mode and the $m=12$ fast mode, respectively.
In both modes, the velocity field is characterized by vortical columns aligned with the rotational ($z$) axis outside the tangent cylinder. The substantial radial motions near the equator indicate that these modes are not toroidal at all. 
The pressure perturbation $p_1$ is in a geostrophic balance with these $z$-vortical motions.
At higher $m$, the fast modes are more strongly confined towards the equator than the slow modes, and their eigenfunctions develop latitudinal sign reversals at middle latitude. 
This behavior arises because their high propagation frequencies give rise to turning surfaces with the presence of strong differential rotation (denoted by gray dashed curves in Fig.~\ref{fig:eig_thRos}d), which confine the mode spatial distribution to the equatorial region \citep[][]{baruteau2013,guenel2016}.

Despite their remarkably similar velocity structures, our eigenmode analysis reveals that the fast and slow modes differ fundamentally in their thermodynamic properties. 
Figures~\ref{fig:thRos_phase}a and b show the equatorial cuts (viewed from the north pole) of the radial velocity $v_r$ and entropy perturbation $s_1$ for the $m=6$ slow and fast modes, respectively. 
It is seen that the longitudinal phases of $v_r$ and $s_1$ are nearly aligned in the fast mode, whereas in the slow mode, the phase of $s_1$ lags that of $v_r$ by approximately $\pi/2m$ in longitude, as quantified by the phase difference shown in Fig.~\ref{fig:thRos_phase}c.
This clearly demonstrates that the fast modes are convective modes that transport thermal energy radially upward. In these modes, buoyancy does positive work to grow the modal kinetic energy rather than serving as the restoring force for the longitudinal propagation.
In this sense, they correspond to the classical thermal Rossby modes, which are intrinsically convectively unstable \citep[][]{busse1970,busse1986b}.
In contrast, the quadrature phase relation in the slow modes implies that buoyancy acts as a restoring force for retrograde propagation. This restoring force nearly cancels the prograde restoring force associated with the compressional $\beta$ effect, thereby substantially reducing the propagation frequency of these modes.
It should be noted that, although buoyancy acts as a restoring force in the slow modes, they should be distinguished from conventional gravito-inertial modes because they exist even in an adiabatic background stratification. 
Since the entropy perturbations in the slow modes originate from thermal diffusion (see Appendix~\ref{appendix:slow}), they should be understood as thermally diffusive modes.

\subsection{High-frequency retrograde (HFR) modes} \label{results:HFR}

In this section, we discuss the high-frequency retrograde (HFR) modes detected by \citet{hanson2022}. 
These modes are characterized by equatorially antisymmetric radial vorticity and propagate retrograde at frequencies about three times higher than those of classical Rossby modes. 
Recent theoretical \citep[][]{triana2022,bhattacharya2023,bekki2024a} and observational \citep[][]{hanson2025} studies have suggested that, unlike the classical Rossby modes, HFR modes are intrinsically non-toroidal inertial modes involving substantial radial motions.

Figure~\ref{fig:eig_HFR}a shows the equatorial power spectrum of north-south antisymmetric component of the latitudinal velocity $v_{\theta}^{-}$ in the middle convection zone.
The red solid curves show the dispersion relation of the HFR modes obtained from the linear model. 
A relatively broad power ridge is distributed near this dispersion curve, which we identify as the HFR modes in the \citetalias{hotta2025} simulation.
In contrast to the prograde columnar modes discussed in Sect.~\ref{results:thRos}, which produce a stronger power concentration in the prograde frequency range, the power of the HFR modes is much weaker and less prominent.
This is also evident in the power spectra at fixed $m$ (Fig.~\ref{fig:eig_HFR}b), where the HFR modes exhibit much broader linewidths and significantly poorer signal-to-noise ratios than the other inertial modes.
Figure~\ref{fig:eig_HFR}c presents the example eigenfunctions of the $m=8$ HFR mode. 
The extracted eigenfunctions are in qualitative agreement with those predicted by the linear model at latitudes lower than the critical layer \citep[e.g., Fig.~6c of][]{bekki2024a}.
In particular, they are confirmed to be non-toroidal modes with substantial radial motions at the equator.

\subsection{High-latitude modes} \label{results:highlat}

High-latitude modes are inertial modes whose power is concentrated at high latitudes and are characterized by spiral flow patterns around the poles. 
Although such flow patterns had long been known observationally \citep[e.g.,][]{ulrich2001,hathaway2013}, they were identified as inertial modes by \citet{gizon2021}.
In particular, the $m=1$ high-latitude mode with north-south symmetric $v_\theta$ (and antisymmetric $v_\phi$) exhibits the largest velocity amplitude among the observed solar inertial modes, reaching approximately $10-20$~m~s$^{-1}$ near the surface.
It has been proposed that the high-latitude modes are excited by baroclinic instability due to the latitudinal temperature gradient in the convection zone \citep[][]{bekki2022a}. 
This theory explains both their observed large amplitudes and the characteristic spiral flow patterns, and has been numerically demonstrated by \citet{bekki2024b} using mean-field simulations that do not explicitly resolve turbulent convection.
Whether the same excitation mechanism operates with the presence of vigorous convection in the \citetalias{hotta2025} simulation is the main focus of the present study.

Figure~\ref{fig:eig_HL}a shows the power spectrum of the latitudinal velocity $v_\theta$ averaged over the northern polar region ($75\degree-90\degree$ in latitude). 
A prominent concentration of power is seen in the retrograde branch at low azimuthal orders, $m \lesssim 4$. 
In fact, this power is a superposition of the high-latitude modes with both north-south symmetries.
To distinguish the contributions from the modes with different symmetries, Fig.~\ref{fig:eig_HL}b shows the 1D power spectra of the north--south symmetric and antisymmetric components, $v_\theta^+$ (solid) and $v_\theta^-$ (dashed), respectively.
First, it is evident that the $m=1$ modes dominate the power spectrum. 
In particular, the $m=1$ mode with the north-south symmetric latitudinal velocity $v_\theta^+$ exhibits the strongest power.
For $m \ge 2$, the power decreases rapidly and the spectral linewidths become progressively broader with increasing $m$, resulting in increasingly poorer signal-to-noise ratios.

Figure~\ref{fig:eig_HL}c shows the spatial eigenfunctions of the most prominent high-latitude mode, namely, the $m=1$ mode with north-south symmetric $v_\theta$ and antisymmetric $v_\phi$.
The horizontal velocity of the mode is concentrated at latitudes above $\sim60^\circ$ and reaches its maximum in a polar-crossing flow with an amplitude of up to $\sim20$~m~s$^{-1}$.
The extracted eigenfunctions exhibit various characteristic features of a baroclinically unstable mode. 
Most notably, a large entropy perturbation is strongly correlated with the latitudinal velocity, indicating that the mode transports thermal energy equatorward (see Sect.~\ref{sec:transport}). 
Figure~\ref{fig:spiral_m1HL} further compares the surface eigenfunctions of the horizontal velocity with those observed \citep[][]{mandowara2026}. 
It is clearly shown that the \citetalias{hotta2025} simulation successfully reproduces the observed spiral flow pattern with the mode amplitude comparable to that observed ($\sim 10$~m~s$^{-1}$) up to $\sim 75\degree$, above which the observational uncertainties increase substantially.
We note that neither the large velocity amplitude nor the spiral structure was reproduced in the earlier simulation of rotating convection \citet{bekki2022b}, where the background baroclinicity was too weak.

\begin{figure}
\begin{center}
\includegraphics[width=0.93\linewidth]{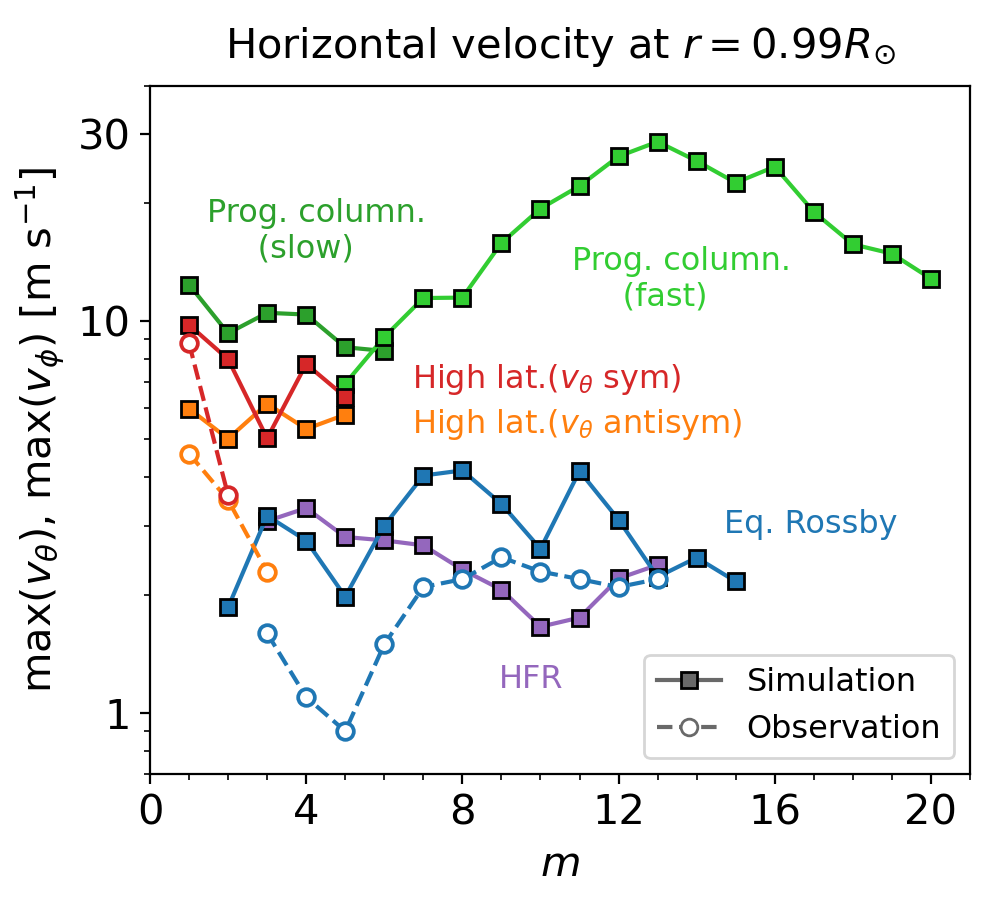}
\caption{
Horizontal velocity amplitudes of the inertial mode at the top boundary ($r=0.99R_{\odot}$).
Maximum latitudinal velocity, max($|v_\theta|$), is plotted for the equatorial Rossby and HFR modes, whereas maximum longitudinal velocity, max($|v_\phi|$), is plotted for the prograde columnar and high-latitude modes.
To ensure a consistent comparison with the observations, the maxima are computed only over latitudes below $80^\circ$.
Open circles with dashed curves denote the observed values reported in \citet{mandowara2026}.
}
\label{fig:vh}
\end{center}
\end{figure}
\begin{figure*}
\begin{center}
\includegraphics[width=0.9\linewidth]{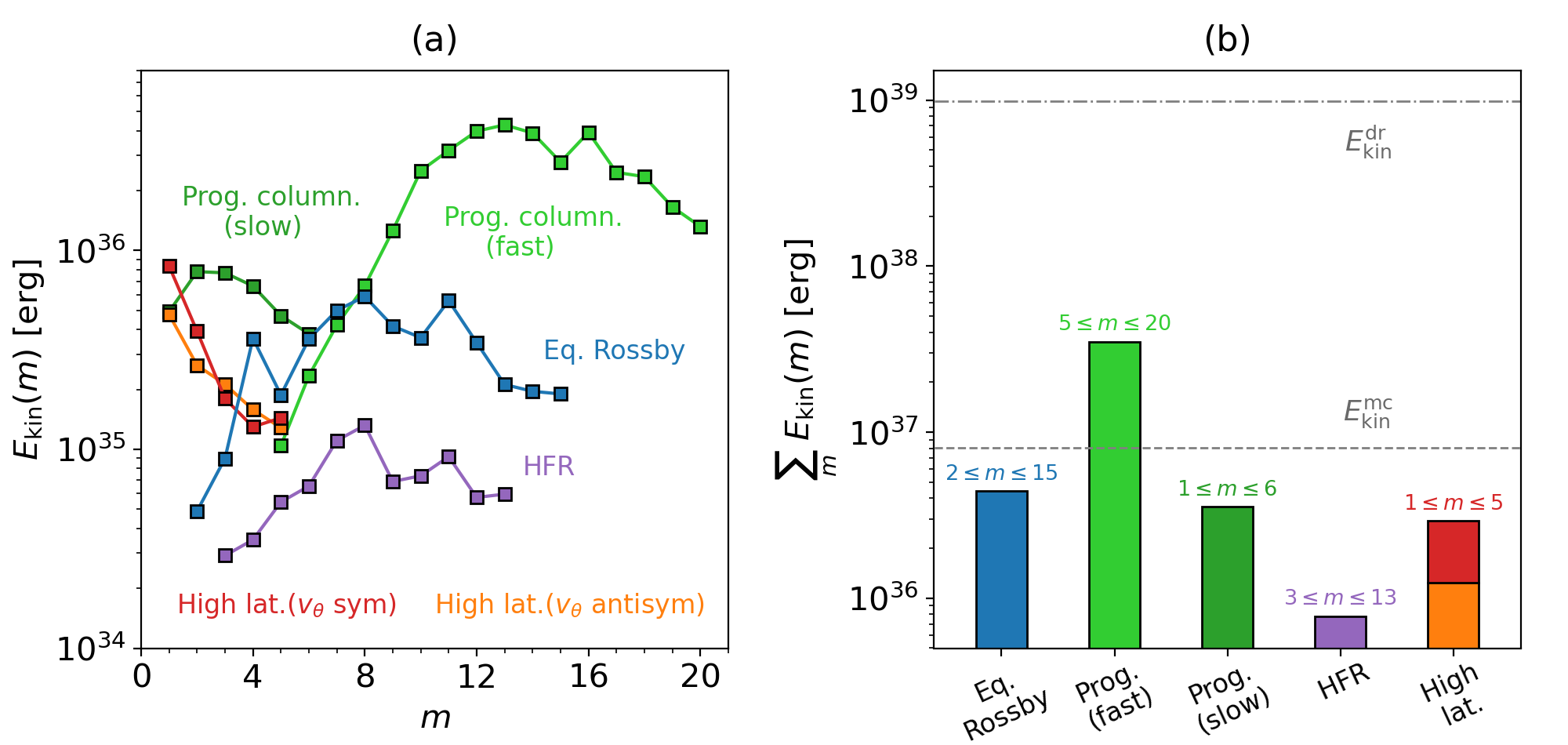}
\caption{
Kinetic energy of the extracted inertial modes.
(a) Mode kinetic energy $E_{\rm kin}(m)$ integrated over the whole convection zone, plotted as functions of $m$. Different colors represent different types of inertial modes.
(b) Total kinetic energy for each type of inertial modes summed over all identified modes.
For comparison, the volume-integrated kinetic energy of the meridional circulation, $E_{\rm kin}^{\rm mc}$, and that of the differential rotation, $E_{\rm kin}^{\rm dr}$, are denoted by gray dashed and dot-dashed lines, respectively.
The range of $m$ taken in each sum is indicated above the corresponding bar.
}
\label{fig:Ekin}
\end{center}
\end{figure*}

\begin{figure*}
\begin{center}
\includegraphics[width=0.75\linewidth]{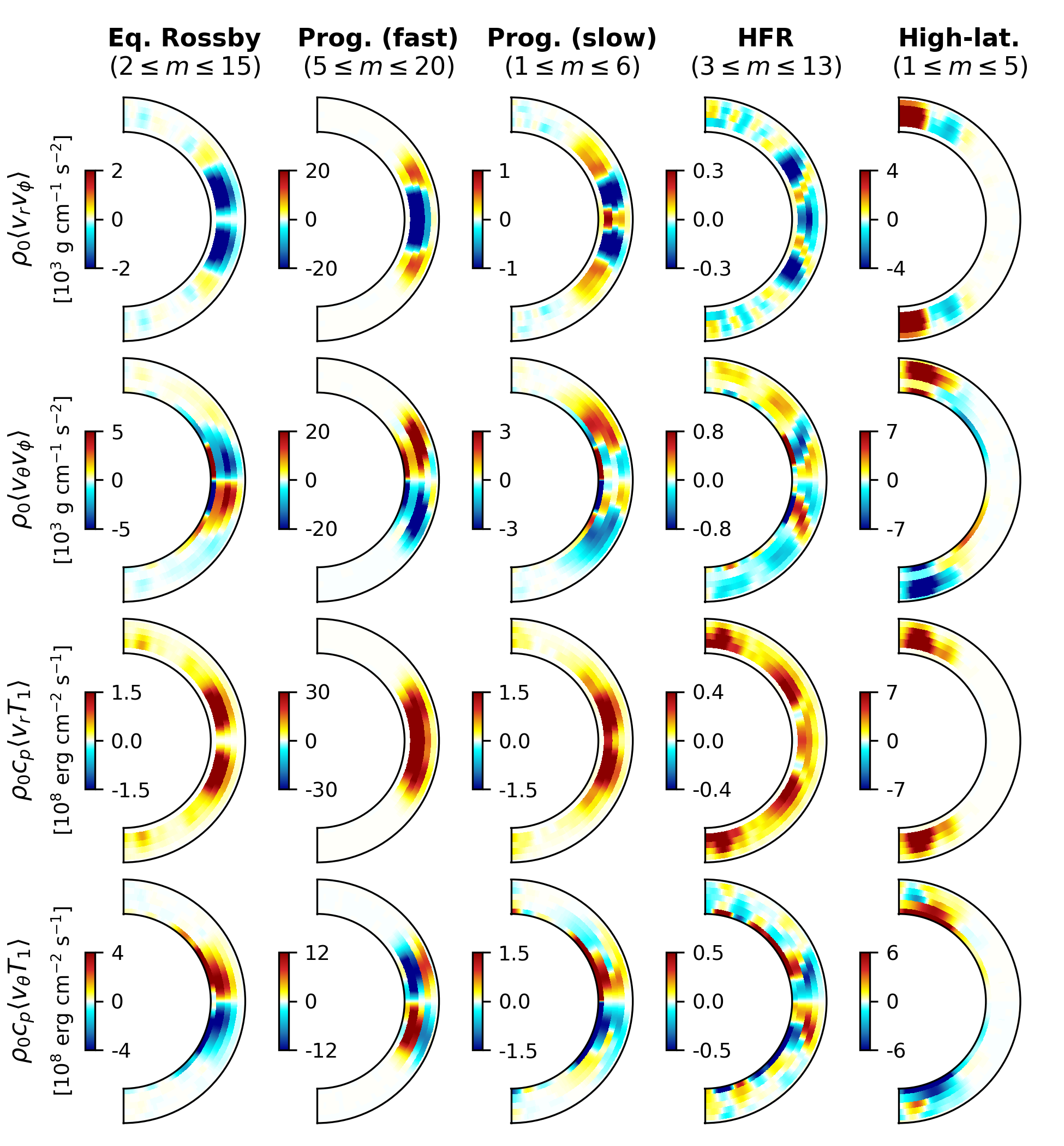}
\caption{
Reynolds stresses and enthalpy fluxes associated with the inertial modes and summed over $m$.
Each column corresponds to a different class of inertial modes -- equatorial Rossby modes, both fast and slow prograde columnar modes, HFR modes, and high-latitude modes with both symmetries -- from left to right.
Top and second rows show the radial and latitudinal components of the Reynolds stress, $\rho_0 \langle v_r v_\phi \rangle$ and $\rho_0 \langle v_\theta v_\phi \rangle$.
Third and bottom rows show the radial and latitudinal components of the enthalpy flux, $\rho_0 c_p \langle v_r T_1 \rangle$ and $\rho_0 c_p \langle v_\theta T_1 \rangle$.
}
\label{fig:RSFe}
\end{center}
\end{figure*}

\begin{figure*}
\begin{center}
\includegraphics[width=0.75\linewidth]{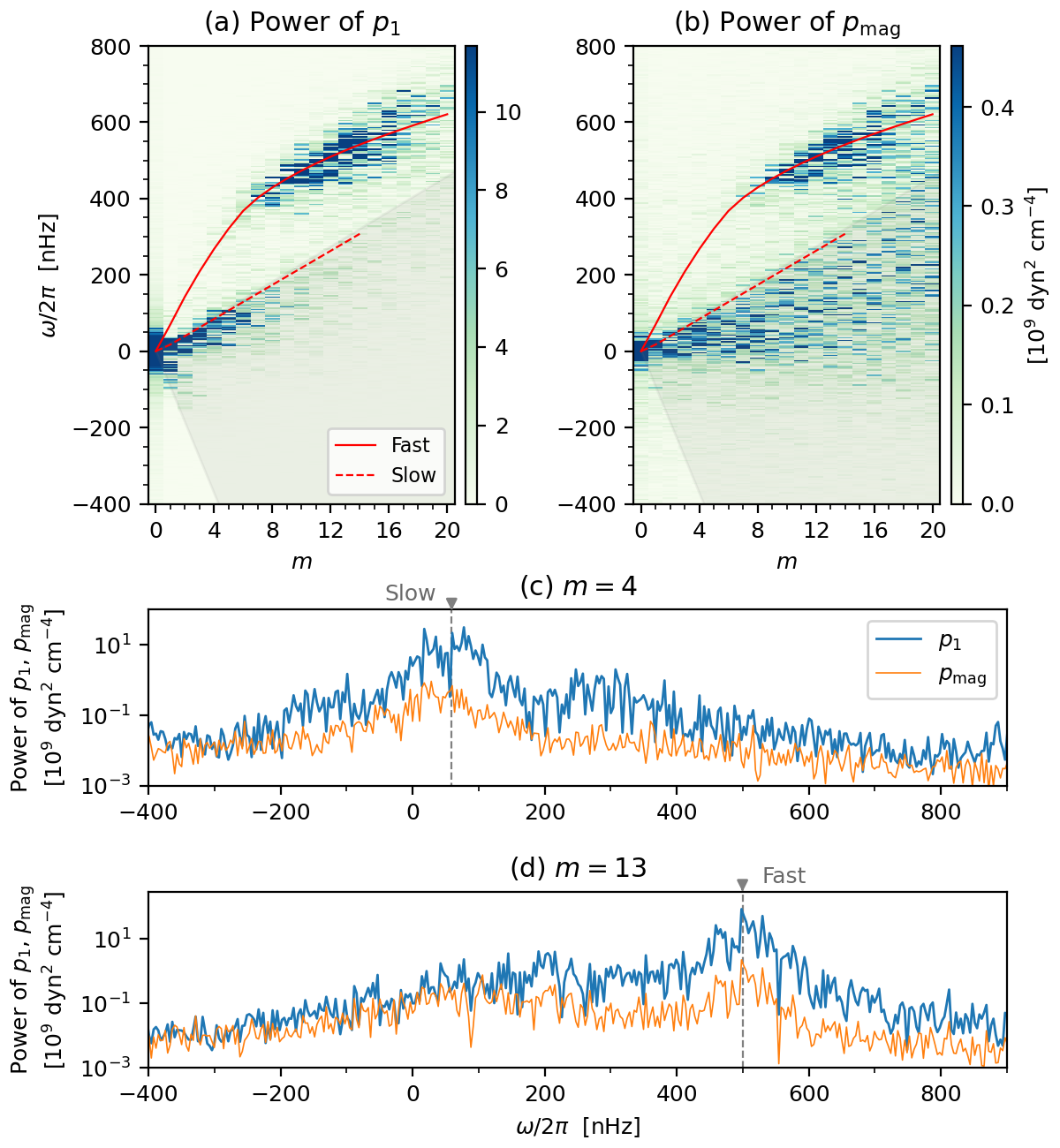}
\caption{
Power spectra of gas pressure perturbation $p_1$ and magnetic pressure $p_{\rm mag}$ near the equator (averaged over $\pm 15^{\circ}$) in the middle convection zone.
Panels (a) and (b) show the $m-\omega$ diagrams of $p_1$ and $p_{\rm mag}$. Red solid and dashed curves show the linear dispersion relations of the prograde columnar modes and low-frequency columnar modes, respectively.
Panels (c) and (d) show the cuts of the spectra at $m=4$ and at $m=13$.
Blue and orange curves show the spectra of $p_1$ and $p_{\rm mag}$, respectively.
}
\label{fig:Power_pmag}
\end{center}
\end{figure*}

\section{Discussion} \label{sec:discussion}

\subsection{Comparison with the observed frequencies} \label{results:comp}

Figure~\ref{fig:disp_comp}a summarizes the dispersion relations of the major inertial modes identified in the \citetalias{hotta2025} simulation in the $m$--$\omega$ diagram, where the frequencies are measured in a frame rotating at $\Omega_0/2\pi = 413$~nHz.
To compare these frequencies with those observed in the Sun, we must first account for the Doppler frequency shifts arising from differences in both the background differential rotation and the reference frame in which the frequencies are measured.
As already discussed in Sect.~\ref{method:H25}, the differential rotation established in the \citetalias{hotta2025} simulation is systematically slower than that of the real Sun. 
We therefore define a reference rotation rate of $\Omega_{\rm ref}/2\pi = 427$~nHz, corresponding to the low-latitude rotation rate in the \citetalias{hotta2025} simulation that is equivalent to the Carrington rotation rate ($\Omega_{\rm Car}/2\pi = 456$~nHz) typically adopted as the reference frame in observations \citep[e.g.,][]{gizon2021}.
We transform the mode frequencies $\omega$ to those seen in the frame rotating at $\Omega_{\rm ref}$ using
\begin{eqnarray}
   &&  \omega_{\rm sim} = \omega + m (\Omega_0 - \Omega_{\rm ref}),
\end{eqnarray}
and plot the normalized frequencies, $\omega_{\rm sim}/\Omega_{\rm ref}$, in Fig.~\ref{fig:disp_comp}b.
For comparison, the observed frequencies measured in the Carrington frame normalized by $\Omega_{\rm Car}$ are overplotted.
It is shown that the frequencies of the equatorial Rossby, HFR, and high-latitude modes present in the \citetalias{hotta2025} simulation are in remarkable agreement with the observations.

In contrast to these robustly observed retrograde inertial modes, the observational evidence for the prograde columnar modes remains controversial, although recent studies have inferred the possible existence of such modes \citep{hanasoge2026, mandal2026}.
In Fig.~\ref{fig:disp_comp}b, the frequencies of the modes interpreted as thermal Rossby modes by \citet[][Table~1; based on measurements of the $t=s-1$ component of the poloidal flow at $0.96\,R_\odot$]{mandal2026} are plotted in lime green for $2 \leq m \leq 18$.
The frequencies of these modes are found to be in excellent agreement with those identified as the slow modes in the \citetalias{hotta2025}  simulation at low azimuthal orders ($m \lesssim 6$). 
This excellent agreement at low $m$ raises the possibility that the modes recently reported by \citet{hanasoge2026} and \citet{mandal2026} may correspond to the thermally diffusive slow prograde columnar modes, rather than to the convectively driven fast modes as typically assumed.
This distinction is important because theoretical studies have long predicted prograde columnar modes as the dominant convectively driven inertial modes \citep[e.g.,][]{miesch2000} and therefore as promising probes of deep convection \citep[e.g.,][]{hanasoge2016}.
If the observed modes belong to the slow branch, previous interpretations relating their amplitudes directly to convective motions may require reconsideration.
Conversely, they may instead carry information about the effective turbulent thermal diffusivity.

Nevertheless, a significant discrepancy remains at high $m$. 
In the \citetalias{hotta2025} simulation, the slow modes become progressively weaker at $m \gtrsim 6$, while the fast prograde columnar modes become exceedingly dominant. 
In contrast, the observations have found the mode power on the slow branch at higher $m$, with no clear evidence for the fast modes. 
Why the classical convectively driven fast modes have so far eluded detection, whereas only the slow modes appear to be observable, remains an open question.
This discrepancy may reflect missing physics in the \citetalias{hotta2025} simulation, limitations of the current observations, or a combination of both, and thus deserves further investigation.

A similar comparison of the mode linewidths with solar observations is presented in Appendix~\ref{appendix:linewidth}.

\subsection{Amplitudes and dynamical impacts} \label{results:amp}

Figure~\ref{fig:vh} shows the maximum horizontal velocity amplitudes of each inertial mode measured at the upper boundary ($r=0.99\,R_\odot$), which provides the most direct quantity for comparison with the observations.
We note, however, that the depths of the simulation and observations differ. The horizontal flows inferred from $5\degree$-tile ring-diagram helioseismology (using f-modes) correspond to a depth of $\sim 2$~Mm below the photosphere \citep{mandowara2026}, whereas the upper boundary of the simulation is located about $\sim 7$~Mm below the photosphere. Furthermore, the impenetrable upper boundary condition ($v_r=0$) might artificially enhance the horizontal velocity amplitudes.

The prograde columnar modes exhibit the largest surface velocities over the entire range of $m$, reaching $10-30$~m~s$^{-1}$. 
This clearly contrasts with the observations, in which no such large-amplitude prograde modes have been detected near the solar surface.
For the high-latitude modes, the $m=1$ mode reproduces the observed amplitude well, whereas the higher-$m$ modes are systematically larger than those observed \citep[][]{gizon2021,mandowara2026}. 
The equatorial Rossby modes exhibit maximum horizontal velocities of approximately $2-4$~m~s$^{-1}$, which are slightly larger than the observed values but remain comparable in magnitude \citep[][]{loeptien2018,gizon2021,mandowara2026}.
A quantitative comparison is not possible for the HFR modes because reliable observational estimates of their amplitudes are currently unavailable. Nevertheless, their amplitudes are known to be systematically smaller than those of the equatorial Rossby modes \citep[][]{hanson2022}, in qualitative agreement with our simulation results.

Figure~\ref{fig:Ekin}a shows the volume-integrated kinetic energy of all identified inertial modes as functions of $m$. 
At $m=1$, the high-latitude mode with north-south symmetric $v_\theta$ has the largest kinetic energy. 
At higher azimuthal orders ($m\ge2$), the prograde columnar modes (both the slow and fast branches) dominate, whereas the HFR modes exhibit the smallest kinetic energies.
Figure~\ref{fig:Ekin}b displays the total kinetic energy summed over $m$ for each inertial-mode family. 
The fast prograde columnar modes are the most energetic, with a total kinetic energy exceeding that of the meridional circulation, $E_{\rm kin}^{\rm mc}$. 
The equatorial Rossby modes, slow prograde columnar modes, and high-latitude modes follow, each having a total kinetic energy less than but comparable to $E_{\rm kin}^{\rm mc}$. 
These inertial-mode families are therefore likely to be dynamically non-negligible in the \citetalias{hotta2025} simulation.
In contrast, the HFR modes possess an order of magnitude less kinetic energy than the other mode families, indicating that their dynamical impact is likely limited.

\subsection{Transport properties} \label{sec:transport}

The upper two rows of Fig.~\ref{fig:RSFe} compare the Reynolds stress components, $\rho_0\langle v_r v_\phi\rangle$ and $\rho_0\langle v_\theta v_\phi\rangle$, summed over $m$ for different inertial-mode families.
The dominant contribution in the equatorial region comes from the fast prograde columnar modes. 
These modes exhibit $\langle v_r v_\phi \rangle < 0$ and $\langle v_\theta v_\phi \rangle > 0 \ (<0)$ in the northern (southern) hemisphere, corresponding to the radially inward and equatorward transport of angular momentum.
We note that, in the \citetalias{hotta2025} simulation, this sign of $\langle v_r v_\phi \rangle$ for the fast prograde columnar modes is opposite to that reported by \citet{bekki2022a,bekki2022b}.
This difference may be related to the background differential rotation:
In those studies, differential rotation was either neglected or remained very weak, under which the prograde columnar modes were found to exhibit $\langle v_r v_\phi \rangle > 0$.
In contrast, in the presence of the strong non-Taylor-Proudman differential rotation realized in the \citetalias{hotta2025} simulation, the fast prograde columnar modes can be substantially modified by the turning surfaces \citep[][see also fig.~\ref{fig:eig_thRos}d]{baruteau2013,guenel2016}, which likely alters their Reynolds stress properties.
A detailed investigation of how differential rotation modifies the Reynolds stresses of prograde columnar modes is left for future work.

It is also seen that the equatorial Rossby modes exhibit weak but predominantly negative (positive) $\langle v_\theta v_\phi \rangle$ throughout the convection zone in the northern (southern) hemisphere. We note that this sign is opposite to that inferred from recent surface measurements by \citet[][]{mandowara2026}, although the origin of this discrepancy remains unclear.

Near the poles, the high-latitude modes predominantly exhibit $\langle v_\theta v_\phi\rangle > 0 \ (<0)$ in the northern (southern) hemisphere, consistent with recent solar surface measurements \citep[][]{mandowara2026}.
Although this Reynolds stress contributes to equatorward angular momentum transport, it has been shown by \citet{bekki2024b} that the net effect of these modes is instead poleward angular momentum transport by the meridional circulation baroclinically driven by their equatorward heat transport.

The lower two rows of Fig.~\ref{fig:RSFe} show the corresponding enthalpy flux components, $\rho_0 c_p \langle v_r T_1\rangle$ and $\rho_0 c_p \langle v_\theta T_1\rangle$.
All inertial-mode families are shown to transport heat radially outward. 
The dominant contribution comes again from the fast prograde columnar modes, indicating that they are convectively driven. 
Their outward enthalpy flux corresponds to $\sim4$\% of the solar luminosity in the equatorial region.
The second-largest contribution arises from the high-latitude modes. Their upward enthalpy flux is consistent with the picture that these modes are excited by baroclinic instability \citep[e.g.,][]{eady1949}.
Regarding the latitudinal enthalpy flux, the fast prograde columnar modes transport heat poleward at low latitudes, whereas the high-latitude modes transport heat equatorward near the poles. 
The latter provides further support for the interpretation that the high-latitude modes are baroclinically excited \citep[][]{bekki2024b}.

\subsection{Effects of turbulent magnetic fields} 
\label{results:pmag}

In this subsection, we examine the influence of magnetic fields on the inertial modes. 
Since the \citetalias{hotta2025} simulation does not generate coherent large-scale magnetic fields, we restrict our discussion to the effects of turbulent magnetic fields.

Since the turbulent dynamo primarily generates mixed-polarity magnetic fields, no clear power ridges associated with the inertial modes are identified in the power spectra of any magnetic-field component, $B_r$, $B_\theta$, and $B_\phi$.
On the other hand, in the power spectrum of the magnetic pressure, $p_{\rm mag}= B^2/8\pi$, we find clear power ridges following the inertial-mode dispersion relations, particularly for the prograde columnar modes.
Figures~\ref{fig:Power_pmag}a and b show the spectra of the north-south symmetric components of the gas pressure, $p_1$, and the magnetic pressure, $p_{\rm mag}$, respectively. 
In both cases, the power is concentrated along the slow and fast branches of the prograde columnar modes, similar to the velocity spectrum (Fig.~\ref{fig:eig_thRos}a).
It is shown, however, that the amplitudes of the magnetic pressure perturbations are more than an order of magnitude smaller than those of the gas pressure (see also the cuts at fixed $m$ in Figs.~\ref{fig:Power_pmag}c,d).
These results indicate that, although the magnetic pressure $p_{\rm mag}$ exhibits low-frequency oscillations corresponding to the inertial modes, the gas pressure $p_1$ remains dominant in the modal dynamics.
In fact, we have also extracted the spatial eigenfunctions of $p_{\rm mag}$ for each inertial mode. 
In all modes, their amplitudes are approximately an order of magnitude smaller than those of $p_1$ (not shown).
These results suggest that, at least in the \citetalias{hotta2025} simulation, the dynamo-generated magnetic pressure does not substantially affect the geostrophic balance and therefore has little dynamical impact on the inertial modes.


\section{Summary and conclusions} \label{sec:summary}

In this study, we analyzed the inertial modes present in the fully nonlinear magneto-convection simulation of \citetalias{hotta2025}, which successfully reproduces the solar-like differential rotation with the solar luminosity and rotation rate in a highly turbulent regime.
Although the \citetalias{hotta2025} simulation does not capture the tachocline, it reproduces remarkably well both the observed departure from the Taylor–Proudman state in the bulk of the convection zone and the NSSL.
Since inertial modes are highly sensitive to the background differential rotation and baroclinicity \citep[][]{bekki2022a,bekki2024b,dey2026}, this simulation dataset provides a unique opportunity to test how well the observed solar inertial modes can be reproduced in a first-principles model of solar-like rotating magneto-convection.

First, we have detected equatorial Rossby modes for $2 \leq m \leq 15$. 
The extracted modes are found to be predominantly the $n=0$ modes with no radial node in the convection zone.
Their frequencies agree remarkably well with solar observations, and their linewidths are also consistent with the observed values (Appendix~\ref{appendix:linewidth}).
Near the surface, they have amplitudes of $2-4$~m~s$^{-1}$, which are only slightly larger than the observed values.

High-latitude modes are also identified at low $m$ ($\leq 5$), with frequencies and linewidths broadly consistent with the solar observations.
Furthermore, thanks to the strongly baroclinic background with a pole-to-equator temperature difference of $-7.5$~K, the simulation reproduces the observed near-surface velocity amplitude of $\sim 10$~m~s$^{-1}$ for the $m=1$ high-latitude mode.
The simulated mode also nicely reproduces the observed spiral flow structure around the poles.
These characteristic features have not been reproduced in previous nonlinear models of rotating convection \citep[e.g.,][]{bekki2022b,blume2024}, in which the differential rotation remained close to the Taylor-Proudman state and the background baroclinicity was much weaker than in the Sun.
These results therefore support the picture that high-latitude modes are baroclinically excited \citep[][]{bekki2022a,bekki2024b}. 
We also note that this instability can arise even in a model without coupling to the underlying radiative interior.

We have also detected HFR modes in the \citetalias{hotta2025} simulation, with frequencies in good agreement with solar observations. 
Compared with the equatorial Rossby and high-latitude modes, the simulated HFR modes have less distinct spectral peaks with broader linewidths and lower power. 
Their low signal-to-noise ratios are qualitatively consistent with those in solar observations \citep[][]{hanson2022}.

In contrast, a discrepancy remains between the simulation and the observations regarding the prograde columnar modes. 
While their existence has recently been observationally inferred using helioseismic normal-mode coupling \citep[][]{hanasoge2026,mandal2026}, they are not robustly detected at the solar near-surface measurements. 
In the \citetalias{hotta2025} simulation, however, prograde columnar modes emerge as the dominant flow pattern throughout the convection zone with their horizontal velocity amplitudes of up to $10-30$~m~s$^{-1}$ near the surface.
Interestingly, the identified prograde columnar modes can be classified into two branches, separated at around $m \simeq 6$.
For $6 \leq m \leq 20$, we identify the classical prograde columnar modes that have been extensively discussed in previous studies \citep[e.g.,][]{glatzmaier1981,bekki2022a,hindman2022}. 
They propagate in a prograde direction at relatively high frequencies comparable to the solar rotation rate, and therefore are referred to as the fast modes.
On the other hand, for $m \leq 6$, we find a distinct class of columnar modes that are almost non-propagating when seen in a frame co-rotating at the equatorial rotation rate. 
We refer to them as slow modes in this study.
Although they possess velocity eigenfunctions remarkably similar to those of the fast modes, their entropy perturbations $s_1$ show a fundamentally different phase relation. 
In the fast modes, the entropy perturbation $s_1$ is nearly in phase with the radial velocity $v_r$, indicating that they are convectively driven \citep[][]{bekki2022b}. 
For this reason, they correspond to the so-called ``thermal Rossby'' modes \citep[e.g.,][]{busse1970}.
In contrast, the slow modes exhibit a quadrature phase relationship between $s_1$ and $v_r$, implying that they are thermally diffusive modes, in which the buoyancy and Coriolis forces nearly cancel with each other and consequently the net restoring force is substantially diminished.
Such thermally diffusive modes have so far received little attention in previous theoretical studies. 
Remarkably, the modes tentatively claimed as thermal Rossby modes by \citet[][]{hanasoge2026} and \citet{mandal2026} have frequencies that agree much more closely with those of the thermally diffusive slow modes than with those of the convectively unstable fast modes.
The reason why the fast prograde columnar modes are so prominent in the simulation at higher $m$, yet remain undetected in the Sun, remains to be understood.

As shown in Sect.~\ref{sec:transport}, the fast prograde columnar modes at high $m$ transport angular momentum radially inward, contrary to previous studies.
Conventionally, a low-Ro regime has been considered favorable for maintaining solar-like differential rotation, as it promotes columnar convection that can transport angular momentum radially outward \citep[e.g.,][]{bekki2022a,camisassa2022,kapyla2023}. 
In the \citetalias{hotta2025} simulation, however, the turbulent magnetic fields can provide the required angular momentum transport, relaxing this constraint on convective flows. 
This finding suggests that the conventional understanding of differential rotation based on Ro regimes and angular momentum transport by columnar convection may need to be reconsidered \citep[][]{hotta2026}.

In the \citetalias{hotta2025} simulation, the turbulent magnetic fields play an essential role. Specifically, they suppress convective motions through the Lorentz force and enhance the efficiency of convective heat transport \citep[][]{rempel2014,hotta2015}, thereby making the background stratification less superadiabatic \citep[][]{bekki2017b,karak2018} as well as enhancing the background baroclinicity \citep[][]{hotta2018}. 
They are also responsible for generating and sustaining the solar-like differential rotation through angular momentum transport by the Maxwell stress \citep[][]{karak2015,hotta2022,bekki2025}.
In this way, magnetic fields are known to have significant impacts on inertial modes, albeit indirectly, by modifying the background profiles, such as the turbulent viscous and thermal diffusivities ($\nu_{\rm t}$, $\kappa_{\rm t}$), the superadiabaticity $\delta$, the latitudinal entropy gradient $\partial s/\partial\theta$, and the differential rotation $\Omega$.
In this study, we have further examined the more direct effect of the magnetic pressure generated by the turbulent dynamo on the geostrophic balance of the inertial modes. 
We analyzed the magnetic pressure perturbation $p_{\rm mag}$ associated with the inertial modes, and found that its power is distributed in spectral space following the dispersion relations of the prograde columnar modes.
However, an examination of the extracted eigenfunctions of $p_{\rm mag}$ reveals that they are an order of magnitude smaller than those of the gas pressure perturbation $p_{1}$.
Therefore, it is implied that the effect of magnetic pressure on the total pressure is not dynamically dominant, at least in the simulation reported here.
This is also suggested by the fact that a hydrodynamic linear model, which does not explicitly include magnetic pressure perturbations, nevertheless reproduces fundamental properties of the inertial modes in the simulation reasonably well.
In other words, provided that the magnetically modified background state is properly taken into account, the traditional hydrodynamic framework \citep[e.g.,][]{bekki2026_review} remains a useful tool for understanding solar inertial modes.

In this study, we used a linear model to aid mode identification (Sect.~\ref{sec:linear}), under several simplifying assumptions such as an adiabatic background ($\delta=0$) and the FOSA approximation for the turbulent viscous and thermal diffusivities (Eq.~\ref{eq:fosa}).
In the \citetalias{hotta2025} simulation, however, the background stratification is non-adiabatic (Appendix~\ref{appendix:delta}), and the turbulent magnetic fields may contribute to the effective viscosity experienced by the modes. 
The simulation datasets, together with the inertial modes extracted in this study, will provide a valuable benchmark for testing methods of inertial-mode helioseismology that use linear models to infer the superadiabaticity and effective viscosity \citep[e.g.,][]{bekki2026_review}.

Finally, we recall that the mechanism responsible for maintaining the Sun's differential rotation remains an open question. Although \citet[][]{hotta2022} proposed that angular momentum transport by turbulent magnetic fields (the so-called punching-ball effect) plays a central role, this scenario has not yet been reproduced independently by other first-principles simulations \citep[e.g.,][]{kapyla2023_review,noraz2025}. 
The present study demonstrates that a simulation operating in such a magnetically dominated regime is simultaneously consistent with a broad range of recently observed properties of solar inertial modes, including several features that have not been reproduced in previous convection-dominated simulations \citep[][]{bekki2022b,blume2024,fuentes2026}. 
The ability to reproduce the observed properties of low-frequency inertial modes, in addition to large-scale mean flows, will thus serve as an important benchmark for assessing how faithfully global simulations reproduce the dynamics of the solar convection zone.


\begin{acknowledgements}
\textit{Author contributions}: YB carried out the analysis (incl. linear calculations), produced all the figures, and wrote the initial draft of the manuscript. HH performed the \citetalias{hotta2025} simulation with \texttt{R2D2} code and provided the data. LG contributed to the project design.
All authors contributed to the final manuscript.  
We thank the International Space Science Institute (ISSI) for supporting discussions within the international team ``Global Inertial Modes: Bridging Solar and Stellar Physics'' (Team 25-637).
YB and LG acknowledge support from ERC Synergy Grant WHOLESUN 810218.
YB's visits to ISEE were supported by the ISEE-MPS joint laboratory project. 
HH further acknowledges support from JSPS KAKENHI (grant No. JP23H01210, JP25H00673, JP25K22031), JST FOREST Program (grant No. JPMJFR246F), and MEXT Next-Generation Computational Science Global Reach \& Innovation Program Japan Grant Number JPMXP1920260401.
The \citetalias{hotta2025} simulation data were generated using the Supercomputer Fugaku provided by the RIKEN Center for Computational Science (hp220173, hp230204, hp230201, hp240219, hp240212, hp250226 and hp250223).
\end{acknowledgements}

\bibliographystyle{aa} 
\bibliography{ref} 

\clearpage

\begin{appendix} 


\section{Superadiabaticity} \label{appendix:delta}

\begin{figure}
\begin{center}
\includegraphics[width=0.85\linewidth]{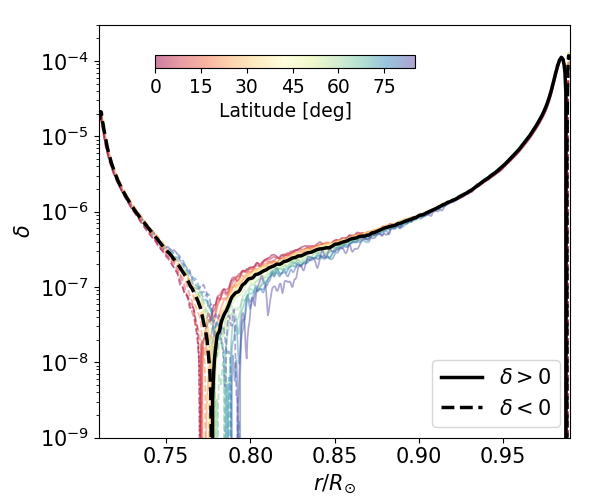}
\caption{
Radial profile of superadiabaticity $\delta$ in the \citetalias{hotta2025} simulation.
Colors indicate latitude, ranging from the equator (red) to the poles (blue). 
The thick black curve shows the latitudinally averaged profile. 
Solid and dashed lines indicate superadiabatic ($\delta>0$) and subadiabatic ($\delta<0$) stratification, respectively.
}
\label{fig:delta}
\end{center}
\end{figure}

In this appendix, we present the superadiabaticity profile realized in the \citetalias{hotta2025} simulation. The superadiabaticity $\delta$ is defined as
\begin{eqnarray}
    && \delta = - \frac{H_p}{c_p} \frac{d\bar{s}_1}{dr},
\end{eqnarray}
where the bar $\bar{ \cdot }$ denotes the horizontal average, and $c_p$ is the specific heat at constant pressure.
As shown in Fig.~\ref{fig:delta}, the mean entropy stratification is weakly subadiabatic near the base of the convection zone, weakly superadiabatic in the bulk, and strongly superadiabatic near the surface.
However, in the linear model employed in this study (Sect.~\ref{sec:linear}), we adopted $\delta=0$ for simplicity, because using the simulated $\delta$ profile directly leads to strong contamination of the inertial-mode spectrum by convectively unstable modes, making the mode identification difficult. 
Despite this simplification, the resulting linear model reproduces the dispersion relations of most inertial modes identified in the \citetalias{hotta2025} simulation reasonably well. 
The small discrepancy for the HFR modes (Fig.~\ref{fig:eig_HFR}a) can be attributed to the absence of the weakly subadiabatic stratification near the base of the convection zone, as discussed by \citet{bekki2024a}.

\section{Physical nature of the slow prograde columnar modes} \label{appendix:slow}

\begin{figure}
\begin{center}
\includegraphics[width=0.96\linewidth]{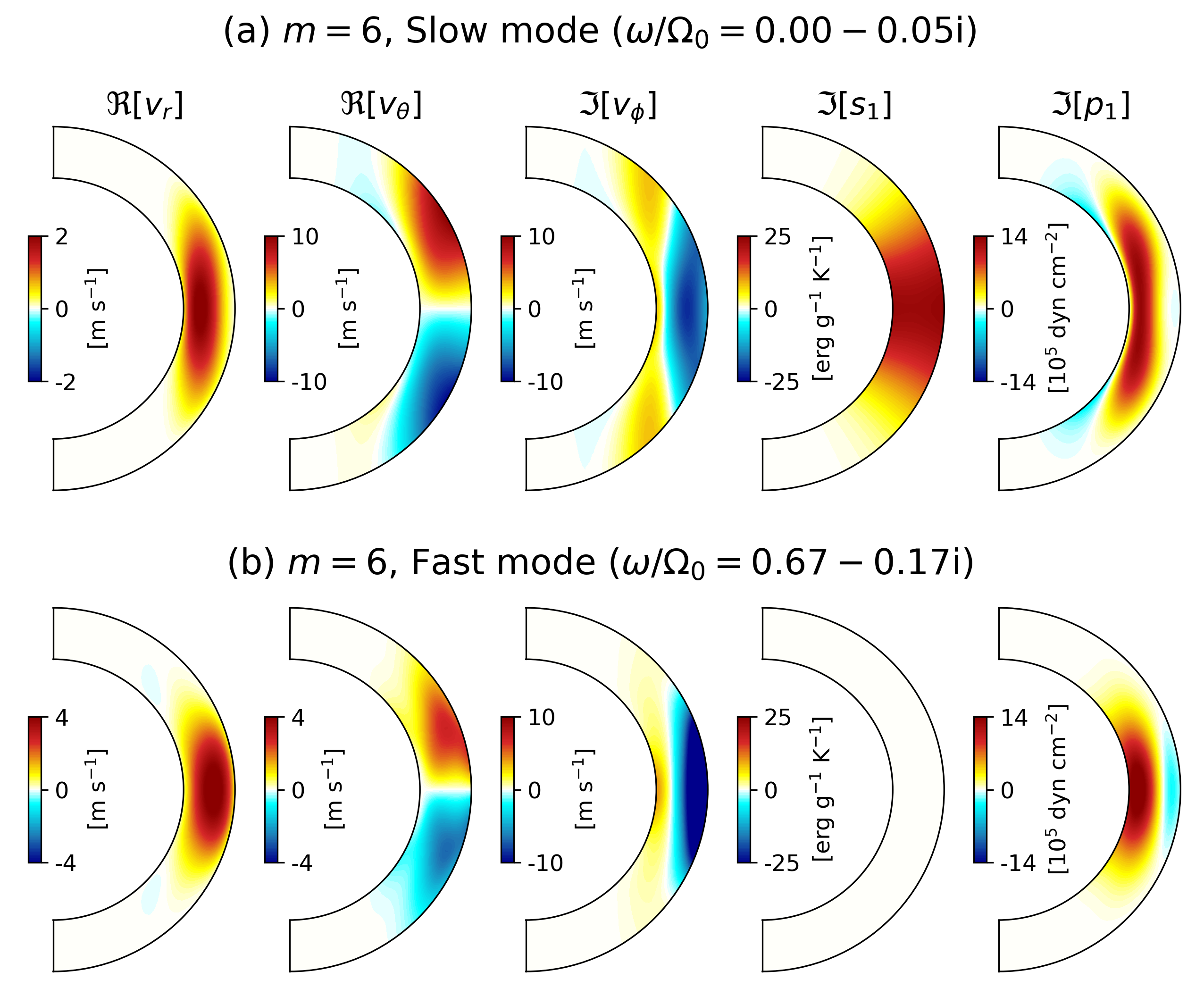}
\caption{
Eigenfunctions of the prograde columnar modes at $m=6$ from the linear model of \citet{bekki2022a} under uniform rotation ($\Omega=\Omega_0$).
The model setup is identical to that described in Sect.~\ref{sec:linear} except that the background radial and latitudinal entropy gradients are set to be zero for simplicity.
(a) Thermally diffusive slow mode with $\Re[\omega]=0$.
(b) Fast prograde columnar mode with $\Re[\omega]>0$ and $s_1=0$.
}
\label{fig:lineig_slow}
\end{center}
\end{figure}

\begingroup
\renewcommand{\thefigure}{C.1}
\begin{figure*}
\begin{center}
\includegraphics[width=\linewidth]{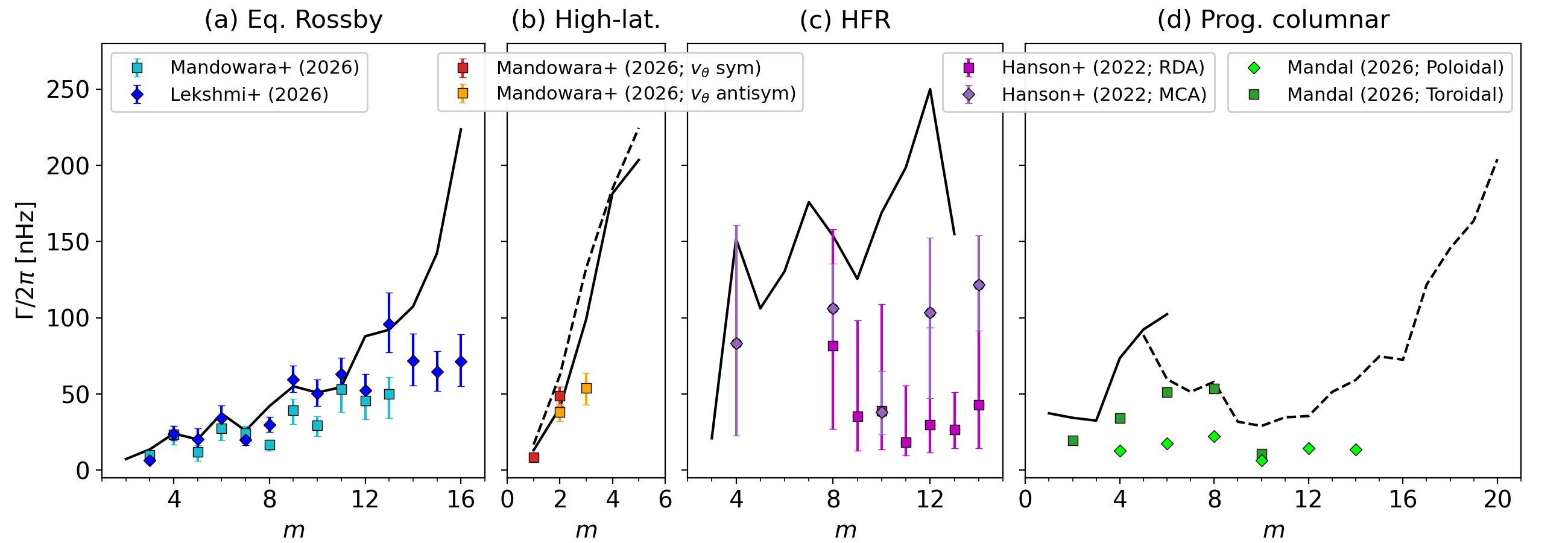}
\caption{
Mode linewidths, $\Gamma/2\pi$, as a function of azimuthal order $m$. 
Panels a--d show the results for equatorial Rossby modes, high-latitude modes, HFR modes, and prograde columnar modes. 
Black curves show the linewidths measured from the \citetalias{hotta2025} simulation (see Eq.~\ref{eq:lorentzian}). 
In panel b, solid and dashed lines correspond to the high-latitude modes with north-south symmetric and antisymmetric $v_{\theta}$.
In panel d, solid and dashed lines correspond to the slow and fast modes, respectively.
Colored symbols show observational values with error bars where available.
The observational results are taken from \citet[][for equatorial Rossby and high-latitude modes]{mandowara2026}, \citet[][for equatorial Rossby modes]{lekshmi2026}, \citet[][for HFR modes]{hanson2022}, and \citet[][for prograde columnar modes]{mandal2026}. 
}
\label{fig:linewidth}
\end{center}
\end{figure*}
\endgroup

In this appendix, we discuss the physical nature of the slow prograde columnar modes found in the \citetalias{hotta2025} simulation at $m \lesssim 6$. 
Since these modes also emerge in the linear analysis as intrinsically thermally diffusive modes, we begin by considering the linearized entropy equation 
\begin{eqnarray}
    && \left[ \omega-m(\Omega-\Omega_0) \right] s_1 = \frac{\ii}{\rho_0 T_0} \nabla\cdot(\rho_0 T_0 \kappa_{\rm t} \nabla s_1). \label{eq:lin_s1}
\end{eqnarray}
Here, we neglect the effects of the background entropy gradient, which are not essential to the following discussion.
This Eq.~(\ref{eq:lin_s1}) admits adiabatic solutions with $s_1=0$ and $\omega \neq 0$. 
These correspond to propagating inertial modes, whose eigenfrequencies are determined by the linearized momentum equation. 
The retrograde-propagating equatorial Rossby and HFR modes, as well as the classical prograde columnar modes (i.e., fast modes) belong to this class.
In the limit of uniform rotation ($\Omega=\Omega_0$), above Eq.~(\ref{eq:lin_s1}) admits another class of solutions with $s_1 \neq 0$ and purely imaginary eigenfrequencies ($\Re[\omega]=0$). 
They are non-propagating modes that simply decay through thermal diffusion, but received little attention in previous theoretical studies so far.

Figure~\ref{fig:lineig_slow} shows the linear eigenfunctions of such non-propagating modes under uniform rotation, compared with those of the prograde mode at $m=6$.
The velocity and pressure eigenfunctions are remarkably similar between the two modes: In both cases, the flow is characterized by $z$-vortical columns outside the tangent cylinder, with the associated pressure perturbation in geostrophic balance with these vortical motions.
The essential difference lies in the entropy perturbation $s_1$. 
In the non-propagating mode, non-zero $s_1$ exists in a phase $\pi/2m$ behind that of the radial velocity $v_r$ in longitude, i.e., $\Re(v_r)$ and $\Im(s_1)$ have the same sign. 
This phase relation implies that the buoyancy force does no net work, but rather acts as a retrograde restoring force that balances the prograde restoring force arising from the compressional $\beta$ effect \citep[][]{bekki2022a,bekki2025}. 
Consequently, the two restoring forces cancel out, yielding a zero propagation frequency.
This phase relation is indeed confirmed in the \citetalias{hotta2025} simulation (Fig.~\ref{fig:thRos_phase}a).

When differential rotation is present, the propagation frequency becomes non-zero due to Doppler shifts, but the underlying physical nature of the thermally diffusive modes remains almost unchanged. 
Indeed, our linear analysis incorporating the differential rotation profile from the \citetalias{hotta2025} simulation successfully reproduces the dispersion relation of the low-frequency columnar modes found in the simulation. 
As shown by the red dashed curve in Fig.~\ref{fig:eig_thRos}a, their frequencies remain very close to zero in a frame co-rotating at the equatorial rotation rate.

So far, the discussion has assumed an adiabatic background stratification ($\delta=0$). 
If the effects of non-adiabatic background stratification are taken into account, the buoyancy force associated with the additional entropy perturbation generated by radial velocity gives rise to an additional restoring force, fundamentally altering the nature of the modes \citep[][Sect.~5]{bekki2022b}.
More specifically, when the background stratification is weakly superadiabatic ($\delta > 0$), the frequencies of the fast prograde modes decrease, whereas those of the slow modes increase, causing the two branches to approach each other. 
At sufficiently large $m$, the two modes merge into a pair of complex-conjugate solutions with the same real frequency, representing growing and decaying modes, respectively  \citep[][]{hindman2023}. 
In practice, only the convectively unstable growing solutions are dynamically important when this happens.

\section{Mode linewidths} \label{appendix:linewidth}

In this Appendix, we compare the mode linewidths, $\Gamma$, obtained from the Lorentzian fitting (Eq.~\ref{eq:lorentzian}) with solar observations. 
In general, linewidths are related to mode lifetimes and damping rates \citep[e.g.,][]{philidet2023}. 
Unlike the frequencies discussed in Sect.~\ref{results:comp}, the linewidths can strongly depend on the observational datasets and fitting methods, thus requiring caution when making quantitative comparisons with observations.

We first compare the equatorial Rossby and high-latitude modes, both of which have relatively high signal-to-noise ratios in solar observations.
Figure~\ref{fig:linewidth}a shows the linewidths of the equatorial Rossby modes measured in the \citetalias{hotta2025} simulation as a black solid curve. 
For comparison, the observational measurements by \citet[][based on HMI $5\degree$-tile ring-diagram data]{mandowara2026} and \citet[][based on GONG $15\degree$-tile ring-diagram data]{lekshmi2026} are shown by cyan and blue symbols, respectively.
The simulated linewidths are slightly larger than but generally consistent with the observed values for $m\lesssim 10$, whereas they increase sharply and deviate significantly from observations for $m\gtrsim 13$.
This rapid increase in $\Gamma$ with $m$ is qualitatively consistent with the picture that the modes are damped by turbulent diffusion \citep[e.g.,][]{fournier2022,philidet2023}.
The results for the high-latitude modes are also shown in Fig.~\ref{fig:linewidth}b.  
The simulated linewidths increase almost linearly with $m$ for modes with both north-south symmetries. 
They are in reasonable agreement with observations for $m=1$ and $2$, while observational constraints are limited at higher $m$ \citep[][]{mandowara2026}.
Overall, the \citetalias{hotta2025} simulation broadly reproduces not only the frequencies of the equatorial Rossby and high-latitude modes but also their observed linewidths.

Figure~\ref{fig:linewidth}c shows the linewidths of the HFR modes. 
As discussed in Sect.~\ref{results:HFR}, the HFR modes in the \citetalias{hotta2025} simulation have poor signal-to-noise ratios and exhibit broad power spectra (see fig.~\ref{fig:eig_HFR}b).
The estimated linewidths are approximately $100-200$~nHz, which are consistent with the results of \citet[][see Appendix~A]{bekki2024a}.
This suggests that the HFR modes are relatively short-lived in convection simulations.
For comparison, we show the observational measurements reported by \citet{hanson2022}, obtained from HMI data using ring-diagram analysis (RDA; magenta) and normal-mode coupling analysis (MCA; purple) in Fig.~\ref{fig:linewidth}c. 
The simulated linewidths are generally larger than the observed values. However, there are also substantial differences between the RDA and MCA measurements, with particularly large uncertainties in the MCA linewidths.

Finally, Fig.~\ref{fig:linewidth}d shows the results for the prograde columnar modes, with solid and dashed curves representing the slow and fast modes.
The linewidths of the slow and fast modes reach their minima of $\sim 40$~nHz at $m \approx 3$ and $10$, respectively, where the corresponding power peaks are very prominent (see fig.~\ref{fig:eig_thRos}b). 
The linewidths tend to increase with $m$ for $m \ge  4$ in the slow modes and for $m \gtrsim 16$ in the fast modes.
For comparison, we also show the linewidths of the observed prograde modes reported by \citet{mandal2026}, obtained from HMI normal-mode coupling analysis. 
The measurements from the power spectra of the poloidal and toroidal flow components are shown by light green and dark green symbols, respectively.
These observational linewidths are generally smaller than those in the \citetalias{hotta2025} simulation. 
However, \citet{mandal2026} did not report the uncertainties, and differences are also seen between the poloidal and toroidal components. 
Furthermore, unlike the maximum-likelihood fitting used in the present study, \citet{mandal2026} employed least-squares fitting, making it difficult to assess the quantitative agreement between the simulation and observations.


\end{appendix}

\end{document}